\documentclass[prb, twocolumn]{revtex4}
\usepackage{graphicx}
\usepackage{amssymb}
\usepackage[latin1]{inputenc}

\begin{document}

%\draft
%\draft
%\title{Zener Tunneling Spin and Valley  Hall Effect  in Topological Insulators}
%\title{Zener tunneling spin and valley  Hall effects in HgTe quantum wells and graphene multilayers }
\title{Zener tunneling isospin Hall effect in HgTe quantum wells and graphene multilayers }
\author{M. Lasia, E. Prada, and  L. Brey}
\affiliation{Instituto de Ciencia de Materiales de Madrid, (CSIC),
Cantoblanco, 28049 Madrid, Spain}

\date{\today}

% buscar PACS
% 73.63.-b Electronic transport in nanoscale materials and structures
% 61.46.-w Structure of nanoscale materials
% 73.22.-f Electronic structure of nanoscale materials: clusters, nanoparticles, nanotubes, and nanocrystals

\begin{abstract}
A Zener diode is a paradigmatic device in semiconductor-based electronics that consists of a $p$-$n$ junction where an external electric field induces a switching behavior in the current-voltage characteristics. We study Zener tunneling in HgTe quantum wells and graphene multilayers. We find that the tunneling transition probability depends asymmetrically on the parallel momentum of the carriers to the barrier. In HgTe quantum wells the asymmetry is the opposite for each spin, whereas for graphene multilayers it is the opposite for each valley degree of freedom. In both cases, a spin/valley current flowing in the perpendicular direction to the applied field is produced. We relate the origin of this \textit{Zener tunneling spin/valley Hall effect} to the Berry phase acquired by the carriers when they are adiabatically reflected from the gapped region.
\end{abstract}

% pseudospin winding number

\maketitle
\section{\label{sec:intro} Introduction}

\begin{figure} [t]
 \includegraphics[clip,width=8cm]{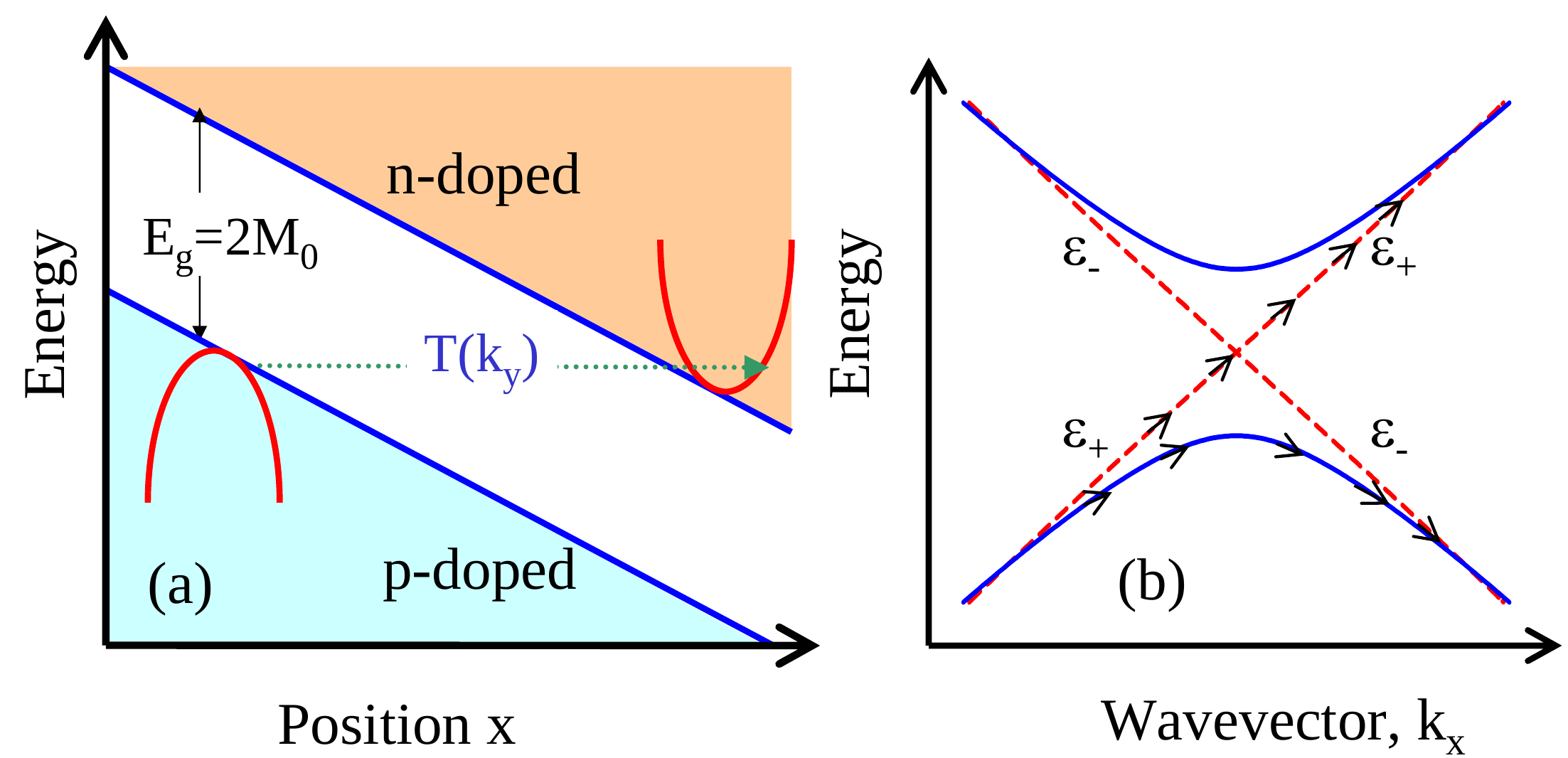}
 \caption{(Color online) (a) Schematic representation of a tunneling process in a Zener diode in the uniform electric field approximation. $T(k_y)$ represents the transition probability of a quasiparticle with momentum $k_y$ from the $p$-doped valence band to the $n$-doped conduction band. (b) Wavevector-energy scheme for the two-band model. The diabatic (adiabatic) energy levels are plotted in dashed red (solid blue) lines. Near the anticrossing region a diabatic $\varepsilon _{+}$ state can tunnel to the diabatic state $\varepsilon _{-}.$
 }
  \label{scheme}
\end{figure}

A large class of semiconductor devices is based on quantum
mechanical tunneling of carriers through potential barriers. This is the case of the \textit{Zener diode}, which consists of a $p$-$n$ junction where a strong enough electric field induces
interband transitions from the valence band of the $p$-type material
to the conduction band of the $n$-doped material, see Fig. \ref{scheme}(a). The tunneling
amplitude is highly non linear in the applied field, and the
tunneling current shows a breakdown-type behavior in the
current-voltage ($I$-$V$) characteristics. The nonlinearity of the Zener
tunneling makes this device very useful for semiconductor-based
electronics\cite{Sze_book}. Interband tunneling has been studied
extensively in parabolic band-gap semiconductors, it is a
paradigmatic example of non-adiabatic transitions, and it is known as
the Landau-Zener tunneling\cite{Zener_1934,Wittig_2005}.
The most used model for studying the  interband tunneling in parabolic semiconductors
is a two-level system described by a Dirac-like Hamiltonian with a
mass term\cite{Kane_Blount,Wittig_2005,Shevchenko_2010},
see Fig. \ref{scheme}(b). In this kind of materials, the spin of the carriers typically plays no role.

In this paper we are interested in analyzing Zener tunneling physics in systems in which there is a correlation between the carrier's spin (or an equivalent degree of freedom) and its direction of motion, i.e., systems is which \textit{chirality} plays a role. In particular, we analyze two types of materials, HgTe quantum wells and carbon-based planar heterostructures. These materials have in common that they can be described by $2\times 2$ Hamiltonians and, therefore, it is possible to map the tunneling problem to the evolution of a two-level system\cite{Shimshoni_1991}.

In the first case, HgTe quantum wells, we find that the Zener tunneling depends asymmetrically
on the parallel momentum of the carriers to the barrier, and this asymmetry is the opposite for each spin.
We call this phenomenon \textit{Zener tunneling spin Hall effect}.
In these quantum wells the central region is an inverted band-gap semiconductor, such as HgTe,
whose intrinsic strong spin-orbit coupling induces an inversion of the normal band progression of
typical semiconductors, like the one used for the barrier material (e.g. CdTe).
This kind of materials have come to the spot-light recently because, depending on the width of the central region,
the system can undergo a quantum phase transition and become a topological
insulator\cite{Qi_2006, Bernevig_2006,Konig_2007}.
A topological insulator is a novel quantum state of matter that has metallic surface states inside the
bulk energy gap \cite{Fu_2007,Moore_2007,Murakami_2007}.
%Note that a spin Hall effect has also been predicted to occur at interfaces between HgTe quantum wells and metals\cite{Guigou_2011}.

%In two
%dimensions the surface states become edge states and the spin Hall
%conductivity is quantized when the Fermi level is inside the
%gap\cite{Qi_2006,Onoda_2005, Bernevig_2006,Konig_2007}.
%Quantized spin Hall effect has been
%predicted to occur\cite{Bernevig_2006} and posteriorly measured in
%mercury telluride quantum wells\cite{Konig_2007}.

In the second case, graphene multilayers, we find that the Zener
tunneling is also asymmetric with respect to the parallel momentum
(except for monolayer), but the asymmetry changes for each carrier's
valley index (instead of real spin). We call this phenomenon
\textit{Zener tunneling valley Hall effect}. Graphitic systems are
also of great interest in condensed matter physics since it became
possible to isolate monolayers, bilayers and in general multilayers
of graphene\cite{Novoselov_2004,Novoselov_2005,Zhang_2005}. $p$-$n$
junctions of graphene have been created by gating locally these
layers and the  transport properties of these heterostructures have
been studied theoretically and
experimentally\cite{Katsnelson_2006a,Cheianov_2006,Williams_2007,Zhang_2008,Young_2009,Vandecasteele_2010,Stander_2009,Jena_2008,Chiu_2010,Brey_2009,Arovas_2010}.
In particular, in bilayer graphene it is possible to open a gap in
the spectrum by applying a voltage difference between the layers,
and Zener tunneling is expected to occur. Actually, it has recently
been predicted that the $I$-$V$ characteristics in bilayer graphene
p-n junctions present, on top of the nonlinear Zener signal, some
N-shaped branches with negative differential
conductivity\cite{Nandkishore_2011}.

In both type of materials, the low-energy Hamiltonian can be
expressed in terms of a \textit{pseudospin vector} that multiplies
the vector of Pauli matrices. In a tunneling process, the pseudospin
vector undergoes a certain trajectory in the Bloch sphere and the
carrier's wave function may acquire a Berry phase. We relate the
Zener transition asymmetry with the spin/valley-dependent Berry
phase that the carriers acquire when they are adiabatically
reflected from the gapped region.

The paper is organized in the following way: In Section II we define the Hamiltonians that govern the properties of HgTe quantum wells and graphene multilayers. In Section III we  map the tunneling problem to the time evolution of a two-level system and show numerical results for the different Hamiltonians. In Section IV we derive analytical expressions for the tunneling transition in the  sudden and adiabatic approximations. In Section V the asymmetry of the tunneling amplitude as function of the momentum parallel to the barrier is explained in terms of the Berry phases that the carriers acquire upon reflection from the barrier. In Section VI we show the $I$-$V$ characteristic curves for HgTe quantum well and multilayer graphene Zener diodes. We finish the paper in Section VII with a summary of our results.

\section{Hamiltonians}

\subsection{HgTe quantum wells}
We study  a HgTe quantum well confined by CdTe
barriers. In bulk, and due to the strong spin-orbit coupling, HgTe
is a zero gap semiconductor. When confined, HgTe is a normal band
insulator for well thickness narrower than 63$\AA$ and becomes a
topological insulator for larger widths\cite{Konig_2007}. For HgTe
quantum wells grown in the (100) direction, the $z$-component of the
spin is conserved. Near the band gap there are four relevant bands:
the E1 bands that consist of the two spin states of the $s$-orbital,
and the two spin states of the HH1 bands which are a linear combination of $p_x$ and $p_y$ orbitals.  The low-energy effective Hamiltonian for the two spin orientations, $s_z
= \pm $1, reads
\begin{equation}
H_{s_z} ({\bf k})= \epsilon (k) \underline I + M(k) \sigma _z + A (
k_y \sigma _y + s_z k_x \sigma _x), \label{Hamil1}
\end{equation}
where $\bf k $=$(k_x,k_y)$ is the in-plane wavevector of the carriers, $k=|{\bf k}|$,
$\epsilon(k)$=$C-D(k_x ^2 + k_y^2)$, $M(k)=M_0+B(k_x^2+k_y^2)$, $
\sigma_x$, $\sigma _y$, and $\sigma _z $ are the Pauli matrices and
$\underline I$ is the identity. $E_g=2M_0$ is the band gap and
$A$-$D$ are parameters fitting the HgTe quantum
wells\cite{parameters}. The product $M_0 B$ determines the character
of the insulator. For $M_0B>0$ the system is a normal insulator,
whereas for $M_0B<0$ a band inversion occurs and the system becomes a
topological insulator. The $p$-$n$ structure of the Zener diode is described by adding to the Hamiltonian the appropriate  scalar external potential of the
form $V(x) \underline I$.

\subsection{Multilayer Graphene}

In graphene and its multilayers, the low
energy properties occur near two non-equivalent valleys ${\bf K}$
and ${\bf K}'$ and the motion of the carriers depends  on the valley
where they reside. The role that the spin plays in HgTe quantum
wells is  played here by the valley index, $\tau _z \pm 1$. Recently, it has been predicted that
spin-orbit coupling in graphene opens a gap and the system could become a topological
insulator\cite{Kane_2005a,Kane_2005b}. However, this gap is very small and the occurrence of the
quantized spin Hall effect would be observed at extremely low
temperatures and in extremely clean samples\cite{Prada_2011}.
%In monolayer graphene, the absence of a gap and
%the chiral nature of the
%carriers makes the tunneling probabilities  larger than in conventional
%semiconductors. In particular, carriers with velocities parallel to
%the electric field, tunnel the barrier with amplitude unity without
%being backscattered.
%In bilayer graphene it is possible to open a gap in the spectrum by applying a voltage difference between the layers, and Zener tunneling is expected to occur.
Thus, neglecting spin-orbit coupling, the low energy properties of $N$-layer
ABC-stacked multilayers are described, in general,  by the
Hamiltonian\cite{Castro_Neto_RMP,Min_2008}
\begin{equation}
H _{\tau_z} ^N = \frac {(v_F p) ^N}{(-\gamma _1)^{N-1}} \left [ \cos (N \phi_{\bf p}) \sigma _x + \sin ( N \phi _{\bf p}) \sigma _y \right ] +
M _0 \sigma _z.
\label{Hamil_N}
\end{equation}
Here the notation is $\cos \phi _{\bf p} =  p _x /p $ and $\sin \phi
_{\bf p} =  \tau _z p_y/p$, where $p_{x,y}=\hbar k_{x,y}$. In the previous expression the Pauli
matrices act on the external layers for $N \ge 1$ and on atoms $A$
and $B$ of the unit cell in monolayer graphene.  $v_F \sim 1
\times$10$^6$ms$^{-1}$ is the velocity of the carriers in monolayer
graphene\cite{Castro_Neto_RMP} and $\gamma _1 \sim 0.3$eV is the
strongest direct interlayer hopping\cite{McCann_SSC}. The last term
in Eq. (\ref{Hamil_N})  opens a gap in the spectrum. In multilayer
graphene this term represents an externally controlled potential
shift in the chemical potential  between the external
layers. In monolayer  graphene, though,  it is not possible
to open a gap experimentally,  but we are going to study this possibility for  the
sake of completeness. Note, however, that at the surface of a
3D topological insulator there will exit a Dirac-like electron
system\cite{hasan_2010,Qi_2011} that, doped with magnetic impurities,
will develop a gap. The band structure of this surface state is
governed by the  same 2$\times$2 Hamiltonian than  monolayer
graphene, but with the $\sigma$-matrices referring to the real electron
spin.

As in the HgTe case, we describe the $p$-$n$ structure of the Zener
diode adding a scalar term, $V(x) \underline I$, to the Hamiltonian
of Eq. (\ref{Hamil_N}).

\section{Constant field and the two-level system}
In this work we
describe Zener tunneling in the uniform electric field model,
$V(x)=- F x $, see Fig. \ref{scheme}(a). In this approximation, it is
possible to get the transmission across the $p$-$n$ junction by
mapping the problem into the evolution of a two-level system\cite{Kane_Blount,Nandkishore_2011}. The key is that for an uniform electric field, $F$, applied  in the $\hat x$-direction, the problem can be simplified if we use the momentum representation, $x=i \partial
_{k_x}$. With it, the Schr\"{o}dinger equation corresponding to
Eqs.  (\ref{Hamil1}) and (\ref{Hamil_N}) becomes
%\begin{eqnarray}
%\! \! \! \! \! \! & & i \hbar F   \frac {\partial \psi _{s_z} }
%{\partial k_x} = \tilde{H} _{s_z} \psi _{s_z} = \nonumber
%\\ \! \! \! \! \! \! & &  \left [ (\epsilon (k) -E) \underline I + M(k)\sigma _z + A (k_y
%\sigma _y+ s_z k_x \sigma _x ) \right ] \psi _{s_z},
%\label{Hamil_2L}
%\end{eqnarray}
\begin{eqnarray}
i F   \frac {\partial \psi _{s} }
{\partial k_x} = [H _{s}({\bf k})-E \underline I] \psi _{s},
\label{Hamil_2L}
\end{eqnarray}
where $E$ is the energy and the index $s$ stands for the spin or the valley index, depending on the system at hand. For each index, $s= \pm 1$,
this equation is identical to the Bloch equation describing the
dynamics of  a spin-$1/2$ particle in the presence of a magnetic field,
with the wavevector in the $x$-direction playing the role of
time. In the uniform electric field model, the term $[\epsilon (k) -E] \underline I$ for HgTe quantum wells in Eq. (\ref{Hamil_2L}), or equivalently the term $E \underline I$ for multilayers, does not contribute to the interband transition and we drop it.

%
%Starting  at $x\rightarrow -\infty$ from the low energy eigenvector
%%of a particle in the valence band
%with $k_x \rightarrow -\infty$ (and eigenvalue $\sigma_{\nu}=-1$)
%and tuning $k_x$ from $- \infty$ to $+ \infty$ as the particles
%crosses to the conduction band at $x\rightarrow\infty$ (and
%$\sigma_{\nu}=1$), the two-level system traverses a level
%anticrossing\cite{Shimshoni_1991}. We have solved numerically Eq.
%(\ref{Hamil_2L}) in  an interval $k_{x,min}<k_x<k_{x,max}$ such
%that, at $k_{x,max}$ and $k_{x,min}$, the eigenvalues of  $\sigma
%_{\nu}$ are $\pm 1$.
%%-1 and 1 respectively.
%From the evolution of Eq. (\ref{Hamil_2L}) we obtain the wavefunction
%at $k_{x,max}$ and from the square of its projection on the state with $\sigma_{\nu}=1$
%we obtain the interband transition probability.

In the limit $k_x \rightarrow \pm \infty$ the eigenvalues of $H_s$
are also eigenstates of $\sigma _{\nu}$, being  $\nu=x$ for
multilayer graphene and $\nu=z$ for HgTe quantum wells. Starting  at
$k_x \rightarrow - \infty$ from the low energy eigenvector (with
eigenvalue $\sigma_{\nu}=-1$) and tuning $k_x$ from $- \infty$ to $+
\infty$, the two level system traverses a level
anticrossing\cite{Shimshoni_1991}. Valence to conduction interband
transitions are described by the process in which a state that at
$k_x = -\infty$ has negative energy evolves into a state that at
$k_x = +\infty$ has positive energy. We have solved numerically Eq.
(\ref{Hamil_2L}) in  an interval $k_{x,min}<k_x<k_{x,max}$ such
that, at $k_{x,max}$ and $k_{x,min}$, the eigenvalues of  $\sigma
_{\nu}$ are $\pm 1$.
%-1 and 1 respectively.
From the evolution of Eq. (\ref{Hamil_2L}) we obtain the
wavefunction at $k_{x,max}$ and from the square of its projection on
the state with $\sigma_{\nu}=1$ we obtain the interband transition
probability.

\begin{figure}
\includegraphics[clip,width=8.cm]{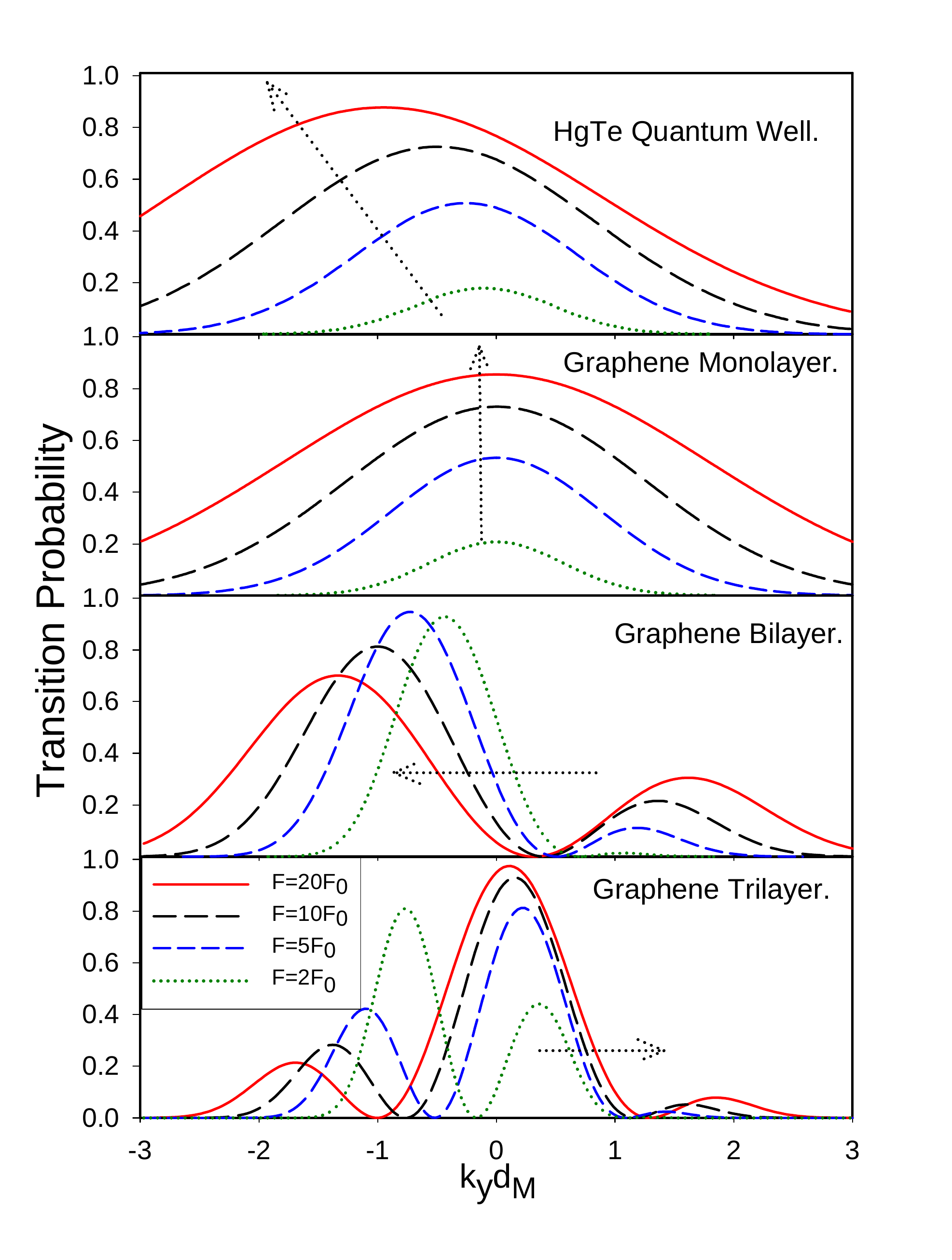}
\caption{(Color online) Zener tunneling in the constant field model as a function of the parallel momentum of the incident particle for HgTe quantum wells and multilayer graphene. The results correspond to spin/valley $s$=1 (for $s=-1$ equivalent results are obtained, but specularly reflected with respect to $k_y=0$). The direction of the arrows indicate the evolution of the curves when increasing  the  electric field.}
\label{Tvsky}
\end{figure}

In Fig. \ref{Tvsky} we plot the Zener transition probability
$T(k_y,s=1,F)$ as function of the wavevector $k_y$ of the incident
particle for a HgTe quantum well and for monolayer, bilayer and
trilayer graphene. We plot the transition probability for several
values of the electric field, $F$=2, 5, 10  and 20$F_0$, with
$F_0\equiv M_0 / d _M$. Here   $F_0$ and $d_M$  are the electric
field and length  characteristic scales set by the gap of the
insulator: $d_M=A/M_0$ for HgTe quantum wells and $d_M$=$ \hbar v_F
( \frac 1 {M_0} \frac 1 {\gamma _1^{N-1}})^{1/N}$ for multilayer
graphene.

The symmetry of the Hamiltonian dictates that
$T(k_y,s,F)$=$T(-k_y,-s,F)$. Except for monolayer graphene, the
transition probability has a maximum at a finite value of $k_y$ that
depends  on the sign of $s$. As a result, carriers with positive
spin/isospin are mainly deflected towards one $\hat y$-direction
when tunneling, whereas those with negative spin/isospin are
deflected in the opposite direction. The overall transition
probability increases with the applied electric field, see Fig.
\ref{Tvsky}. At small fields the spatial extension of the forbidden
region becomes very large and the Zener tunneling amplitude goes to
zero abruptly when $F \rightarrow 0$. This is the origin of the
switching behavior of the Zener diodes. For moderate applied
electric fields the asymmetry in the angle of incidence also
increases with the field.

\section{Sudden and Adiabatic approximations}
In order to shed some light on our numerical results, we have solved Eq. (\ref{Hamil_2L}) analytically in the limit of small parallel momentum
$k_y$ and  large electric field. The analytical calculations expand
the solution of the Hamiltonian in a {\it diabatic} or in an {\it
adiabatic} basis, see Fig. \ref{scheme}(b). The first case is suitable for an unperturbed Hamiltonian that can be diagonalized in a diabatic basis, where the carriers evolve with probability one from the
valence to the conduction band.
We then
calculate the first correction to perfect transmission
in the sudden approximation, treating  the
rest of the Hamiltonian in first order perturbation theory. When the Hamiltonian is such that the
tunneling transmission form valence to conduction band is very small, it is more convenient to use the
adiabatic basis as the unperturbed one. In the adiabatic
basis the carriers are completely  reflected at the barrier, and the
tunneling probability can be obtained as first order perturbation to
the adiabatic Hamiltonian.

The Hamiltonians of the systems we are studying can be written in
the form
\begin{equation}
H= \varepsilon ({\bf k})\, \vec{h} ({\bf k}) \cdot \vec{\sigma}. \label{twobytwo}
\end{equation}
This equation defines a wavevector-dependent unitary pseudospin vector
$\vec{ h} ({\bf k})$.
%The Chern number corresponding to the
%Hamiltonian Eq,\ref{twobytwo} is the number of times the unit sphere
%is covered by pseudospin when ${\bf k}$ runs over the whole
%reciprocal space\cite{Hasan_2010,Qi_2010}.
%\begin{equation}
%n= \frac 1 {4 \pi} \int d ^2 k \left ( \partial _{k_x}
%\overrightarrow{h} \times
%\partial _{k_ y} \overrightarrow{h} \right ) \cdot\overrightarrow{ h} \, \, \, .
%\end{equation}
From the form of this Hamiltonian, the expectation  value of the vector of
$\sigma$ matrices is either parallel, in the conduction band, or antiparallel, in the valence band, to the pseudospin. In the absence of gap,
$M_0=0$, carriers approaching perpendicularly to the barrier, $k_y
=0$, should conserve the pseudospin. In agreement with the Klein
paradox, that implies perfect transmission for gapless  monolayer
and trilayer  graphene and perfect reflection for gapless bilayer
graphene\cite{Katsnelson_2006a}. This conservation of the pseudospin
at $k_y=0$ for massless Hamiltonians would help us to choose a
diabatic or adiabatic basis as the starting point in perturbation
theory.

\subsection{HgTe quantum wells, diabatic basis and sudden approximation}
In HgTe quantum wells, the pseudospin has the form
\begin{equation}
\vec{h}^{HgTe} \! = \! ( Ak_x,A k_y s_z,M_0 \!+ \! B k^2
 )/\sqrt{(M_0 \! + \! Bk^2)^2 \! + \! A^2 k^2}  . \label{h_HgTe}
\end{equation}
For $M_0=0$ and $B=0$, the pseudospin  takes the form: $(k_x,k_ys_z,0)/|k|$,
 and the eigenfunctions  of the Hamiltonian of Eq. (\ref{Hamil1}) are
chiral. In this limit, the Klein paradox dictates\cite{Katsnelson_2006a}
that the tunneling amplitude at $k_y=0$ is unity.  For finite values
of $M_0$ and $B$, the Klein paradox does not apply exactly, but the
transmission probability at large electric fields and small values of
$M_0$ and $k_y$ is close to unity. Therefore, it is convenient to work in
the diabatic basis.

\begin{figure}
 \includegraphics[clip,width=8cm]{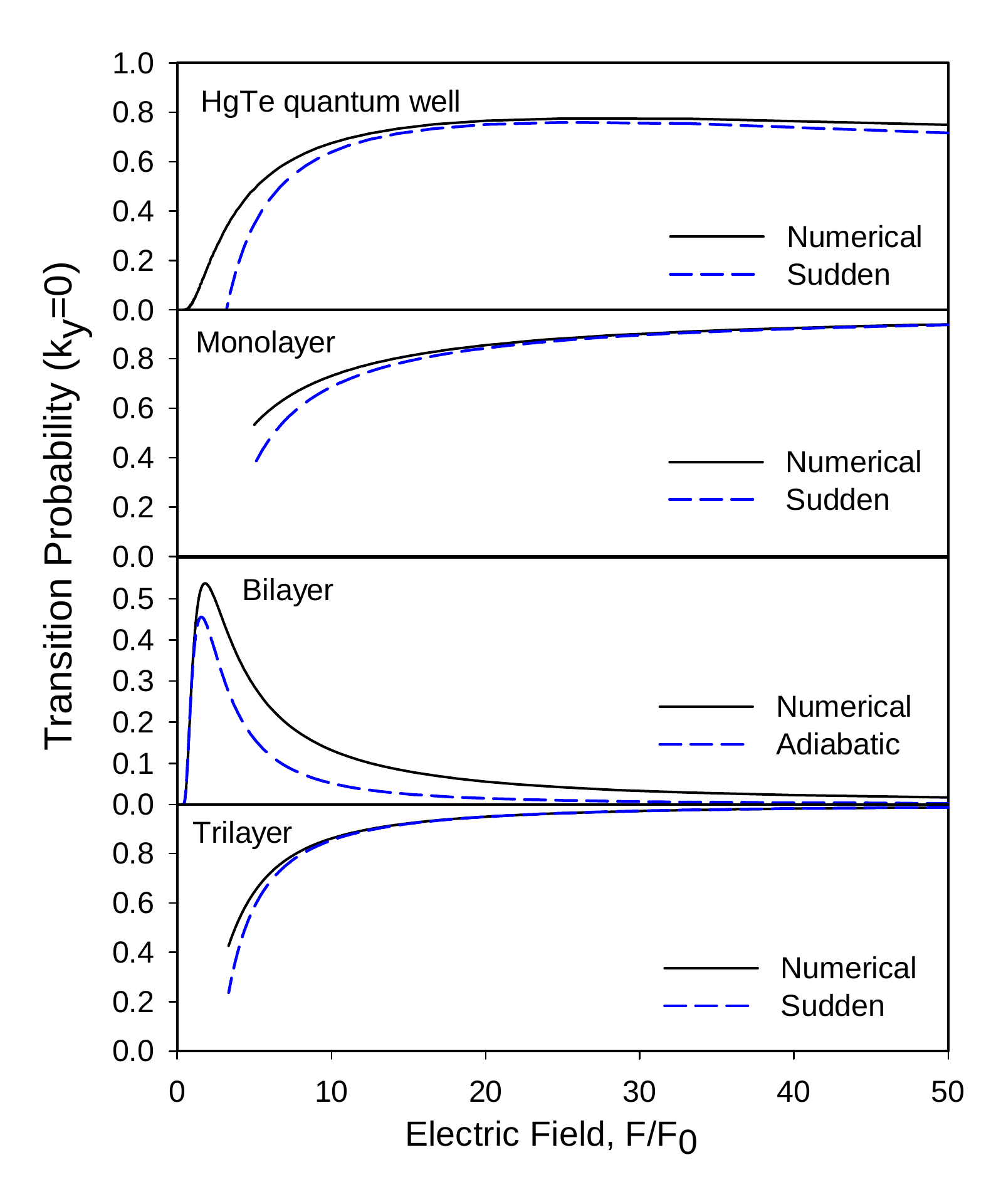}
 \caption{(Color online) Comparison of the transmission probability at $k_y=0$ obtained by solving numerically Eq. (\ref{Hamil_2L}) with the  analytical result obtained in first-order perturbation theory: Eq. (\ref{app_hgte}) for HgTe quantum wells, Eq. (\ref{app_m}) for graphene, Eq.(\ref{app_b}) for bilayer graphene and  Eq. (\ref{app_t}) for trilayer graphene.}
  \label{Appro}
\end{figure}

In the
natural units of the problem, $x \equiv k_x d_M$, $y
\equiv k_y d_M$, $\mathcal{E} \equiv F/F_0$ and $\tilde{B} \equiv
B/(M_0 d _M  ^2)$, the Hamiltonian is written as:
\begin{eqnarray}
& &  \! \! \! \!\!  \! \! \!\!\! i   \frac {\partial }{\partial x} \psi
\equiv (H_0 + \tilde{V}) \psi \, \, \,
, \textrm{with} \, \, \nonumber \\
& &  \! \!\!\! \! \! \! \! \! \! H_0 \! = \!\frac 1  {\mathcal{E}} \! \left(
  \begin{array}{cc}
    \!\tilde{B} x ^2 &  \!x s _z \\
    \!x s _z  & \!-\tilde{B} x ^2 \\
  \end{array}
\!\!\right), \,  \tilde{V} \!= \!\frac 1 {\mathcal{E}}\!
\left(
  \begin{array}{cc}
    \!1\!+\!\tilde{B} y ^2 & \!-i y \\
   \! iy  & \!-1\!-\!\tilde{B} y ^2 \\
  \end{array}
\!\!\right)\!.
\label{Hamil3}
\end{eqnarray}
The eigenvectors of $H_0$ define the diabatic basis, with
eigenvalues $\varepsilon _{\pm}(x) = \pm  \frac x
{\mathcal {E} } \sqrt{ 1+ \tilde{B} ^2 x ^2 }$, see
Fig. \ref{scheme}(b). We represent the corresponding wavefunction in
the diabatic basis as
\begin{equation}
\psi (x) =  C _  -( x) e ^{- i \omega(x)}|->+ C _+( x) e ^{+ i
\omega(x)} |+>, \label{wfHgTe}
\end{equation}
where $H_0 | \pm > = \varepsilon _{\pm} | \pm >$ and $ \omega (x) =
 \int ^ x \varepsilon_{+}(x') dx'$. We define the
transmission ${t}$ and reflection amplitude ${r}$ as follows:
assuming that $C _ - (- \infty)$= 0 and $C _ +(- \infty)$= 1, then
${r}= C_ - ( \infty)$ and ${t}=C_+ (\infty)$. Plugging the
wavefunction, Eq. (\ref{wfHgTe}), into the Hamiltonian of Eq.
(\ref{Hamil3}), we get
\begin{equation}
\partial _x C _{\pm} = \pm \frac  i {\mathcal E} \mathcal{T}_0  C_ {\pm} -
\frac  i {\mathcal E} \mathcal{T} _{\pm}  C_{\mp}, \,  \,  \,
\,\mathrm{ where} \label{edif}
\end{equation}
\begin{eqnarray}
\! \!\! \! \! \! \! \! \! \! \! \! \!  \! \!& &\mathcal{T}_0  \! \! = \! \!
(1 + \tilde B y ^2) \frac { \tilde B x ^2} {\sqrt{ x ^2 + {\tilde B } ^2  x ^4}}  \, \, \, \,  \mathrm{and}  \nonumber \\
\! \! \! \! \! \!\! \! \! \! \! \! \! \! \! & & \mathcal{T} _{\pm}
\! \! = \! \mp  e ^{ \mp 2 i \phi(x)} \! \! \left (  \! \! s_z x \frac { 1 +
\tilde {B} y ^2} {\sqrt{ x ^2 + {\tilde B } ^2 x ^4}} \! \!+\! \! i y \! \! + \! \!i
\mathcal{E} s _z \frac {\tilde {B}/2 }{1 + \tilde B x ^2} \right ).
\end{eqnarray}
The amplitudes $r$ and $t$ are obtained from the
asymptotic solution of Eq. (\ref{edif}) with the appropriate boundary
conditions. For large values of $\mathcal E$ it is possible to get
an analytical expression for the transition. To lowest order in
$1/\mathcal {E}$, the asymptotic  form of $C_  -$ is obtained by
substituting $C_ + (x)=1$ in Eq. (\ref{edif}),
\begin{equation}
C_- (\infty) = - \frac i  {\mathcal E} \int _ {-\infty} ^{+\infty}
\mathcal{T}_{-} dx.
\end{equation}
For small   values of $ {\tilde B}^2 {\mathcal E} $ this integral
can be evaluated using the steepest descent method. From it we
obtain the interband transition probability
%\begin{eqnarray}
% \! \! \! \! \!  & &  T(k_y, s_z) =   1- |\tilde{r}|^2  \simeq \nonumber \\
%\! \! \! \! \! & &1-\frac {\pi}{ F} \frac {M _0 ^2} {A} \left ( 1 +
%s_z k_y \frac {F  B}{M_0 ^2}+ \frac {k_y^2}{M_0 ^2} (A^2+ 2BM_0)
%\right )
%\end{eqnarray}
\begin{eqnarray}
 \! \! \! \! \!   \! \! \! \! \! & &  T(y, s_z) =   1- |r|^2  \simeq \nonumber \\
\! \! \! \! \!  \! \! \! \! \! & &1 \! - \! \frac {\pi}{ \mathcal E}
\! \left (\! \left [ \!1 \! +\! \tilde{B} y ^2 \!+\!\frac {
\tilde{B} ^3 {\mathcal E} ^2} 4 \! \right ] ^2 \!+\!   \left [
y\!+\! \frac {{\mathcal E} \tilde{B}} 2 s_z ( 1 \!- \! \frac
{\tilde{B}} 2 ) \right ]  ^2 \!
 \right ).
 \label{app_hgte}
\end{eqnarray}
Recovering previous units, the maximum of the transition
probability to lowest order in $F B ^2$ occurs at a wavevector
\begin{equation}
k_y ^M = -  s _z  \frac  {F  B} {A^2} \left (1-\frac 5 2 \frac B
{A^2} M_0 \right ), \, \, \, \label{kym}
\end{equation}
being the maximum transition probability
\begin{equation}
T_{max}= 1-\frac {\pi} F \frac   {M_0^2} A \, \, \, . \label{tmax}
\end{equation}
In Fig. \ref{Appro} we compare the transmission at $k_y$=0  obtained numerically from Eq. (\ref{Hamil_2L}) with the one obtained from Eq. (\ref{app_hgte}).
The quality of the approximation is good, specially for strong electric fields.
Equations (\ref{kym}) and
(\ref{tmax}) explain qualitatively much of the results presented in the first panel of
Fig. \ref{Tvsky}:
 i) for each spin orientation the transition
probability is asymmetric with respect to $k_y$,
 ii) the asymmetry increases
with the field, iii) the sign of
$k_y^M$ depends on the product  $s_z B$ (note that when $B=0$ there is no spin Hall effect), iv) the asymmetry is present either for $M_0B>0$ or $M_0B<0$, i.e., irrespective of whether the quantum well is in the trivial or in the topological phase,
and v) the overall transition increases with the
electric field.
Moreover, Eqs. (\ref{kym})-(\ref{tmax}) describe
quantitatively the results in the case of large $F$.
For example, for $F/F_0$=10  we get $T_{max}$=0.69 and
$k_y^M=0.45$, results that are in rather good agreement with the numerical ones presented in Fig. \ref{Tvsky}.

A spin dependent transmission has been also predicted to occur at
the interface between a HgTe quantum well and a
metal\cite{Guigou_2011}. In this case the asymmetry is related to
localized states at the interface.

\subsection{Monolayer graphene, diabatic basis and sudden approximation}
The pseudospin vector for graphene has the form
\begin{equation}
\vec{h}^{m} = (v_F p_x, v_F \tau _z p_y,M_0)/\sqrt{M_0 ^2+v_F ^2
p^2}, \label{h_mono}
\end{equation}
and for  gapless monolayer graphene the Klein paradox applies exactly.
Therefore, in order to study the tunneling when $M_0 \ne 0$, it is
convenient to work in the diabatic basis and, using natural units as before, we write the Bloch-like
equation as
\begin{equation}
i \frac \partial {\partial x} \psi = \left [ \frac 1 { \mathcal E}
\left(
  \begin{array}{cc}
    0 &  x \\
    x  & 0 \\
  \end{array}
\!\!\right)+
\frac 1 { \mathcal E}
\left(
  \begin{array}{cc}
    1 &  -i y \tau_z \\
    iy \tau _z & -1 \\
  \end{array}
\!\!\right) \right ] \psi .
\label{hamil_m}
\end{equation}
Note that in the reduced units, $\hbar v_F$ plays the same role as $A$.  The eigenvalues of the first term of Eq. (\ref{hamil_m}) define the diabatic basis. Using this basis, the wavefunction takes the form
\begin{equation}
\psi (x) =  C _  -( x) \frac 1 {\sqrt{2}}\left (\begin{array}{c}
                                           1 \\ -1
                                         \end{array} \right )
e ^{i  \frac {x ^2}{2{ \mathcal E}}}+
C _  +( x) \frac 1 {\sqrt{2}}\left (\begin{array}{c}
                                           1 \\ 1
                                         \end{array} \right )
e ^{-i  \frac {x ^2}{2{ \mathcal E}}},
\label{wf}
\end{equation}
and the coefficients $C_-$ and $C_+$ satisfy
\begin{equation}
\partial _x C _{\mp}= - \frac i  { \mathcal E} (1-i y \tau _z) e ^{i  \frac {x ^2}{2{ \mathcal E}}} C_{\pm}.
\end{equation}
The reflection amplitude in the sudden approximation, valid in the ${ \mathcal E}  \rightarrow \infty $ limit, is then
\begin{equation}
r \simeq - \frac i { \mathcal E}(1-iy \tau _z) \int _{-\infty} ^{\infty} e ^{i  \frac {x ^2}{2{ \mathcal E}}}dx,
\end{equation}
being the transition probability, in the original units,
\begin{equation}
T(k_y)=1-\frac \pi {F \hbar v_F} (M_0 ^2 + \hbar ^2 k_y ^2 v_F ^2) \, .
\label{app_m}
\end{equation}
In agreement with the numerical results, we get that the transition probability is symmetric in $k_y$ and independent on the isospin $\tau _z$ (see Fig. \ref{Tvsky}).
Note that, for monolayer graphene, the transition probability can be obtained exactly\cite{Kane_Blount,Zhang_2008},
\begin{equation}
T(k_y)= e ^{-\frac \pi {F \hbar v_F} (M_0 ^2 + \hbar ^2 k_y ^2 v_F ^2)},
\end{equation}
and Eq. (\ref{app_m}) corresponds to the first term in the $1/F$ expansion.
Fig. \ref{Appro}, second panel, illustrates the quality of the sudden approximation at large values of the electric field.

\subsection{Bilayer graphene and the adiabatic approximation}
For bilayer graphene, the pseudospin has the form
\begin{equation}
\vec{h}^{b} = \left(\frac {v_F ^2} {\gamma_1} (p_x ^2 -p _y ^2),\frac
{v_F ^2 } {\gamma_1} 2p_x p_y \tau_z ,M_0\right)/\sqrt{M_0^2+\frac {v_F^4}
{\gamma_1 ^2} p^4}. \label{h_bilayer}
\end{equation}
Gapless bilayer graphene, $M_0 =0$, is also chiral, and  the
pseudospin is $(k_x^2-k_y^2,2k_xk_y \tau _z ,0)/k^2$. Holes
impinging perpendicularly to the barrier from the left have opposite
pseudospin than electrons moving to the right, and the same
pseudospin than holes reflecting from the barrier. Therefore, the
tunneling probability for $M_0=0$ and $k_y =0$ is null. When $M_0
\ne 0$, the transition probability at $k_y =0$ is still small, see
Fig. \ref{Tvsky}, and it is thus more appropriate to work in the
adiabatic basis.

The Bloch equation for bilayer graphene has the form
\begin{equation}
-i  { \mathcal E} \frac \partial {\partial x} \psi =  \left(
  \begin{array}{cc}
    1 &   (x-i y \tau_z)^2\\
    (x+i y \tau_z)^2  \tau _z  & -1 \\
  \end{array}
\!\!\right)\psi = H_B \psi \, \, , \label{hamil_b}
\end{equation}
where now $ { \mathcal E} = \frac {\hbar v_F}{M_0} \frac F
{\sqrt{M_0 \gamma_1}}$.

The Hamiltonian $H_B$ in Eq. (\ref{hamil_b}) defines the adiabatic
basis with eigenvalues $\varepsilon (x,y)= \pm \sqrt{1 +
(x^2+y^2)^2}$ and eigenfunctions
\begin{equation}
\psi _- \! \!  = \! \! \left (\begin{array}{c}
                                           - \sin{(\theta /2)} \\ \cos {(\theta /2)} e ^{i \phi}
                                         \end{array} \right )  \, \, {\rm and} \, \,
\psi _+ \! \!  = \! \! \left (\begin{array}{c}
                                           \cos{(\theta /2)} \\ - \sin {(\theta /2)} e ^{i \phi}
                                         \end{array} \right ),
\end{equation}
where $\cos {(\theta)} = 1/|\varepsilon|$ and $\phi = \tan ^{-1} {(\frac {-2xy} {x^2-y^2}} )$ .
In order to solve Eq. (\ref{hamil_b}) we consider the general solution
\begin{equation}
\psi = a_1  e ^{-i \frac {\omega (x,y)} {\mathcal E}} \psi_- + a_2  e ^{i \frac {\omega (x,y)} {\mathcal E}} \psi _+,
\end{equation}
with $\omega (x,y) =  \int _0 ^x \varepsilon (x',y) dx'$. The coefficients $a_1$ and $a_2$ satisfy
\begin{eqnarray}
\frac {\partial} {\partial  x} a_1&  = &  - a_2 {\mathcal T}(x,y) -i a_1  \cos ^2 \left ( \frac \theta 2 \right ) \frac {\partial \phi}{\partial x}
\nonumber \\
\frac {\partial} {\partial  x} a_2&  = &  - a_1 {\mathcal T}^* (x,y) -i a_2  \sin ^2 \left ( \frac \theta 2 \right )   \frac {\partial \phi}{\partial x},
\end{eqnarray}
with
\begin{equation}
{\mathcal T} = e ^{2 i \frac {\omega(x,y)} {\mathcal E}}  \frac {iy(3x^2-y^2)+x(x^2-3y^2)/\varepsilon ( x,y)}{ (x^2 + y^2) \varepsilon ( x,y)}.
\end{equation}
Now we take the adiabatic limit, i.e., we consider that the probability to undergo a transition from one adiabatic state to another is negligible, $a_2 \sim 1$. Then the transmission amplitude is
\begin{equation}
t(y)= \int _{-\infty}  ^{\infty}   {\mathcal T} (x,y)dx.
\label{I_b}
\end{equation}
In the limit $y \rightarrow 0$ this integral can be evaluated following the methods presented in Refs. [\onlinecite{Davis_1976, Shimshoni_1991}] and we obtain (back in physical units), see appendix A,
\begin{eqnarray}
T(k_y, \tau_z)&  \approx   & \frac {4 \pi ^2 } 9
e ^{ -  2 c_1 \frac {M_0}{F} \frac {\sqrt {M_0 \gamma _1}} {\hbar v_F} }\sin ^2  \left (c_1  \frac {M_0}{F} \frac {\sqrt {M_0 \gamma _1}} {\hbar v_F} \right )
\nonumber \\  & \times &
\left ( 1- c_2   k_y \tau _z  \left (\frac F {M_0} \right ) ^{1/3} \frac {(\hbar v _F) ^{4/3}} {(M_0 \gamma_1) ^{2/3}} \right),
\label{app_b}
\end{eqnarray}
where $c_1 \approx 1.23$ and $c_2\approx 3.19$ are numerical factors.
Some comments on Eq. (\ref{app_b}) are in order: i) as shown in Fig. \ref{Appro}, there is a reasonable agreement between the numerical results and the one obtained in the adiabatic approximation, ii) the transition probability is not symmetric with respect to $k_y$, but it is so with respect to the product $k_y \tau _z$, iii) the maximum transition occurs at finite $k_y$, iv) there is an oscillatory term in the transmission amplitude that produces zeros in the tunneling probability at finite values of $F$, $M_0$ and $k_y$. These zeros appear because the bilayer graphene Hamiltonian is quadratic in the momentum and, for each energy in the gap region, there are two  decaying states that interfere  under the tunneling barrier\cite{Nandkishore_2011}.

\subsection{Trilayer graphene, diabatic basis and sudden approximation}
In ABC-stacked trilayer graphene the pseudospin unitary vector is
\begin{equation}
\vec{h}^{t} = \frac {\left( \frac {v_F ^3} {\gamma_1 ^2} (4 p _x ^3 -3
p_x p^2 ),\frac {v_F ^3 } {\gamma_1 ^2 } (3 p^2 p_y -4 p_y^3
)\tau _z,M_0\right)}{ \sqrt{M_0^2+\frac {v_F^6} {\gamma_1 ^4} p^6}}. \label{h_trilayer}
\end{equation}
For massless trilayer graphene the pseudospin  reduces to $(4
k_x^3 -3k_x k^2,(3k_y k^2-4k_y^3)\tau_z,0)/k^3$ and the eigenvectors are
again chiral. Because the pseudospin rotates 6$\pi$ when the
wavevector rotates 2$\pi$ around $\mathbf{k}=0$, the transition
probability at $k_y$=0 is unity for massless trilayer graphene. In
Fig. \ref{Tvsky} we see that, even for $M_0 \ne 0$, in the limit of
large electric field the transition probability at small $k_y$ is
near one. Therefore, it is appropriate to work in the diabatic
basis and use the sudden approximation. We write the Bloch equation
as the sum of a diabatic term plus a perturbation,
\begin{equation}
i \frac \partial {\partial x} \psi  \!  = \! \frac 1 { \mathcal E} \left [
\left( \! \!
  \begin{array}{cc}
    0 &  x^3 \\
    x ^3  & 0 \\
  \end{array}
\!\!\right) \!+\!
\left(\! \!
  \begin{array}{cc}
    1 &   (x \! \! - \!  \! i \tau _z y)^3 \! \! -\! \! x^3\\
   (x \!  \! + \! \! i \tau _z y)^3 \! \! -\! \! x^3& -1 \\
  \end{array}
\!\!\right)  \! \! \right ]  \! \! \psi  ,
\label{hamil_t}
\end{equation}
and in the case of the trilayer graphene we have
\begin{equation}
{\mathcal E} = \frac {M_0}{ F} \frac {(\gamma_1 ^2 M_0)^{1/3}}{\hbar v_F} \, .
\end{equation}
The eigenfunctions of the first term of Eq. (\ref{hamil_t}) define the diabatic basis. In this basis the wavefunction takes the form
\begin{equation}
\psi (x) =   \frac {C _  -( x)} {\sqrt{2}} \left (\begin{array}{c}
                                           1 \\ -1
                                         \end{array} \right )
e ^{i \ \frac {x ^4}{4{ \mathcal E}}}+
\frac {C _  +( x)} {\sqrt{2}}\left (\begin{array}{c}
                                           1 \\ 1
                                         \end{array} \right )
e ^{-i \frac {x ^4}{4{ \mathcal E}}}
\, \, \,
, \label{wf_t}
\end{equation}
and the coefficients $C_-$ and $C_+$ satisfy
\begin{equation}
\partial _x C _{\mp} \! = \!  \mp \frac i { \mathcal E} 3xy^2 C_{\pm} \! + \! \frac i { \mathcal E}   (1 \pm i y \tau _z (y^2 \!  -  \! 3x^2))
e ^{-i \frac {x ^4}{2{ \mathcal E}}} C_{\mp} \, \,.
\end{equation}
To lowest order in $1/{\mathcal E}$ and $y$, the reflection amplitude is
\begin{eqnarray}
\! \! \! r  \! \! \!  & =  & \! \! \! C_-(+ \infty) \simeq - \frac i { \mathcal E} \int _{-\infty} ^{\infty}(1-i 3 x ^2 y \tau _z)  e ^{i  \frac {x ^4}{2} \frac 1{ \mathcal E}}dx \, \, \, \,
 \nonumber \\ & = &  \! \!  \! \left (  \frac  1{\mathcal E} \right ) ^{3/4} \! \!
\frac {\Gamma(1/4)}{2^{3/4}} e ^{i \frac {3 \pi} 8} \!+\! \left (
\frac  1 {\mathcal E} \right ) ^{1/4} \! \! \! \! 3 \tau_z y \frac
{\Gamma(3/4)} {2^{1/4}}e ^{-i \frac {3 \pi} 8} \label{app_tr}
\end{eqnarray}
and the transmission probability, in the physical units, is
\begin{equation}
T(k_y,\tau _z) = 1- \frac {M_0 ^2 \gamma _1}{(\hbar v_F F)^{3/2}} \frac {\Gamma ^2 (1/4)}{2 ^{3/2}} - \tau _z \frac {M_0 k_y} F
\frac {\Gamma (1/4)\Gamma (3/4)}{\sqrt{2}}.
\label{app_t}
\end{equation}

This expression agrees remarkably well with the numerical results for
large electric fields, see Fig. \ref{Appro}. Also, Eq. (\ref{app_t})
explains qualitatively the dependence of the tunneling probability on
the wavevector $k_y$.

\section{Berry phase and lack of reflection symmetry}

The question that remains is the physical origin of the asymmetry,
for a fixed spin/valley,  of the tunneling amplitude as a function
of $k_y$. The asymmetry is not related to Chern number associated
with the chirality of the massless, $M_0=0$,
Hamiltonians\cite{Prada_2011}. Although HgTe and monolayer graphene
share the same Chern number, in monolayer graphene the transition
amplitude is symmetric with respect $k_y$, whereas it is not so in
HgTe quantum wells.

We associate the asymmetry with the winding of the expectation value
of the pseudospin $ \vec{h} ({\bf k})$  when a carrier is adiabatically
reflected by the tunneling barrier. This is related to the sign of the Berry phase acquired by the carrier's pseudospin in this process.

Consider  a quasiparticle  moving in the valence band in the presence of a
constant electric field, $V(x)=-Fx$. This quasihole coming from  $x
=-\infty$ and moving towards the right has a momentum $k_x
<0$. Upon arriving into the gapped region, it is adiabatically reflected from  it back to $x = -\infty$ with momentum  $k_x>0$.
In the presence of the electric field, the momentum $k_x$ is not a good
quantum number and it is not  conserved. In a
semiclassical/adiabatic  approximation the momentum is defined by
the relation
\begin{equation}
\varepsilon ( \tilde{k} _x(x),k_y) -F x = E \, \, .
\end{equation}
In the reflection process the pseudospin $ \vec{h} (\tilde{k}_x ,
k_y)$ describes a trajectory on the Bloch sphere of radius unity. When the
trajectory closes a circuit $\Gamma$ in the unit sphere, the
wavefunction of the carrier acquires a Berry phase equal to half the
solid angle defined by the surface enclosed  by the circuit
$\Gamma$.

\begin{figure}
 \includegraphics[clip,width=8cm]{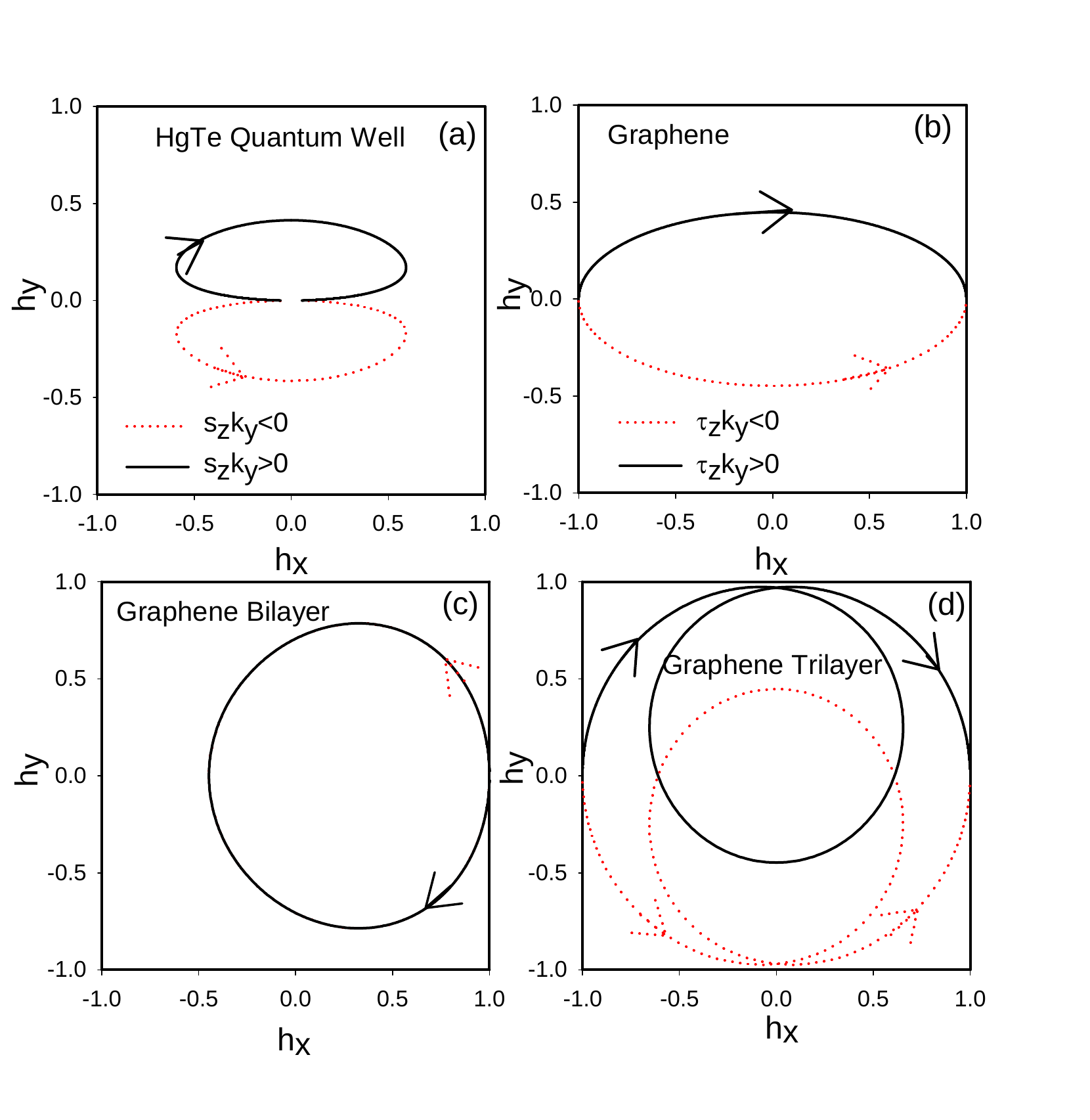}
 \caption{(Color online) In-plane projection of the trajectory defined
 by the pseudospin  $ \vec{h} (\tilde{k}_x ,
k_y)$ in an adiabatic reflection process.  Solid lines correspond to trajectories with $s k_y >0$ and dotted lines to trajectories with
$s k_y <0$. }
  \label{Isospin}
\end{figure}

In Fig. \ref{Isospin} we plot, for the different Hamiltonians studied in
this paper, the in-plane projection of the trajectories defined by the pseudospin $\vec{h}$
when the carrier goes from $x = -\infty$ to the barrier and is
reflected adiabatically back to $x = -\infty$. In the case of
monolayer graphene such trajectory defines an open line, both for $\tau_z k_y$ greater or smaller than zero.
Thus, for monolayer graphene there are no closed paths in the adiabatic
process and there is no Berry phase associated with the
reflection. The situation is different for HgTe quantum
wells. In this case the trajectory defines a closed circuit and there
is a Berry phase associated with the adiabatic reflection. The sign of the Berry phase depends on the direction in which the closed loop is traversed by the pseudospin. It turns out that
it has opposite sign for opposite signs of
$k_y$ or $s_z$. Therefore, the sign of the Berry phase depends on the
sign of the product $k_y s_z$. For graphene multilayers, $N>1$, the
pseudospin trajectory in the adiabatic reflection process always
defines closed paths that have opposite orientation for opposite
values of the product $k_y \tau_z$. The dependence of the Berry
phase on the product $k_y s $, being $s$ the spin or the valley
index, breaks the reflection symmetry in each index $s$  and explains
the asymmetry of the transmission for a momentum $k_y$ at a fixed
index $s$.

\begin{figure}
 \includegraphics[clip,width=8cm]{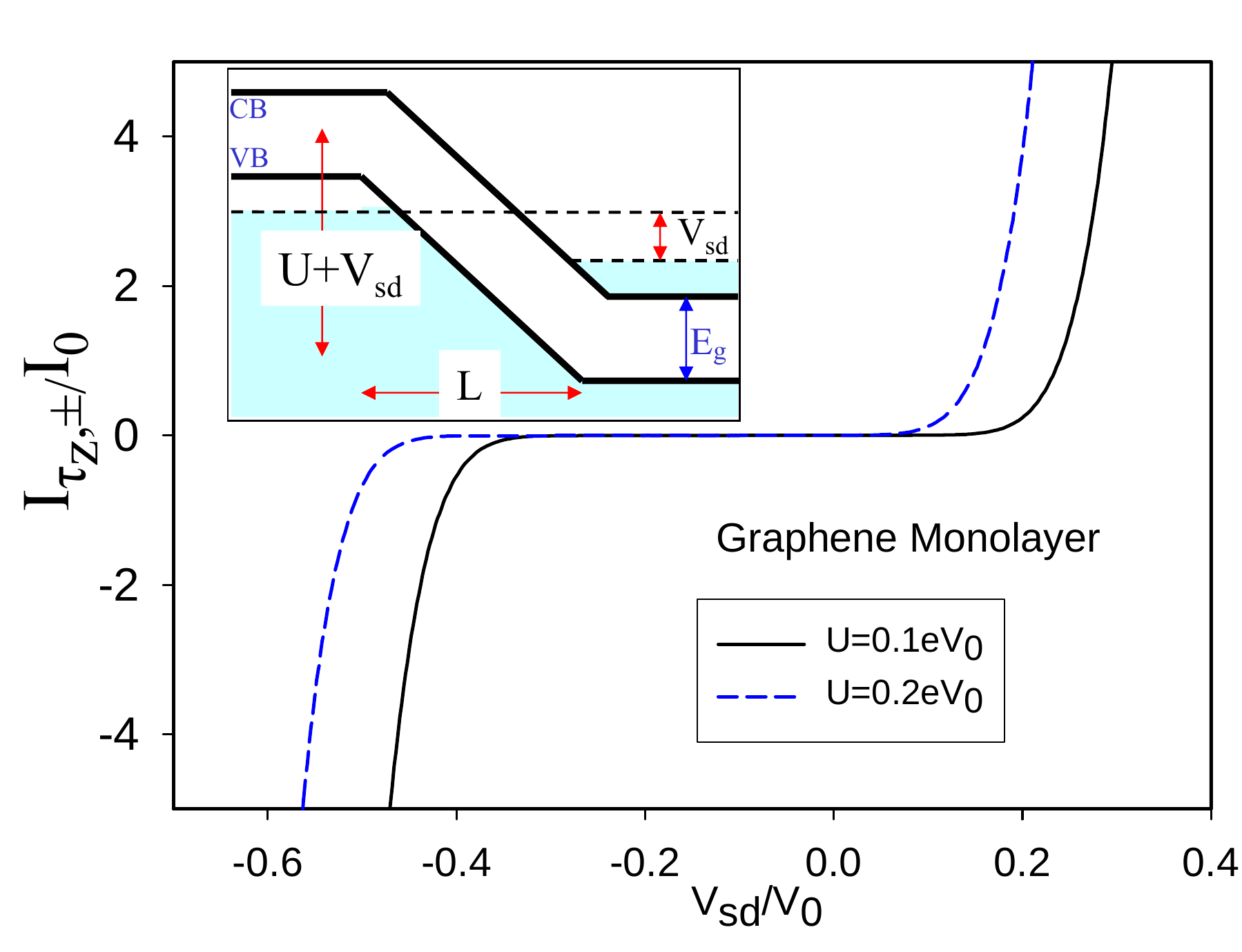}
 \caption{(Color online) I$_{\tau_z, \pm}$-V characteristics of monolayer graphene
 for two values of the built-in potential. Units are $I_0= \frac {e^2}h V_0 W/d_M \times 10^6$ and $eV_0 = M_0 L/d_M$.
 The inset shows schematically the Zener diode.}
  \label{I_Vsd_monolayer}
\end{figure}

\begin{figure}
 \includegraphics[clip,width=8cm]{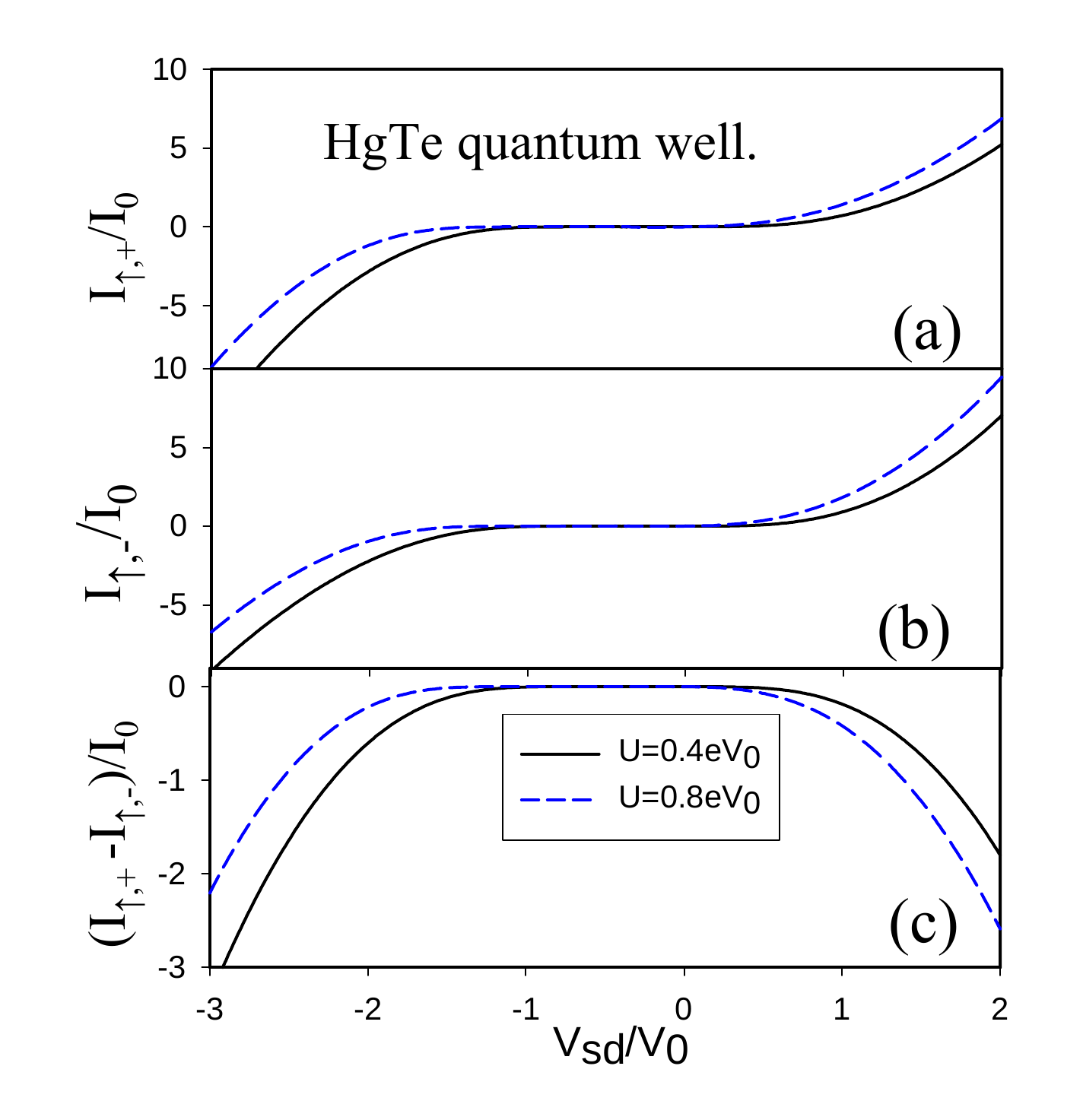}
 \caption{(Color online) I$_{\uparrow, \pm}$-V characteristics of HgTe quantum
 wells,
 for two values of the built-in potential. Panels (a) and (b) correspond to current flowing in the
 positive and negative
 $\hat y$-direction, respectively. In panel (c) we plot the excess of current in the positive
 $\hat y$-direction for electrons with spin up. Units are $I_0= \frac {e^2}h V_0 W/d_M \times 10^2$ and $eV_0 = M_0 L/d_M$.}
  \label{I_Vsd_TI}
\end{figure}

\begin{figure}
 \includegraphics[clip,width=8cm]{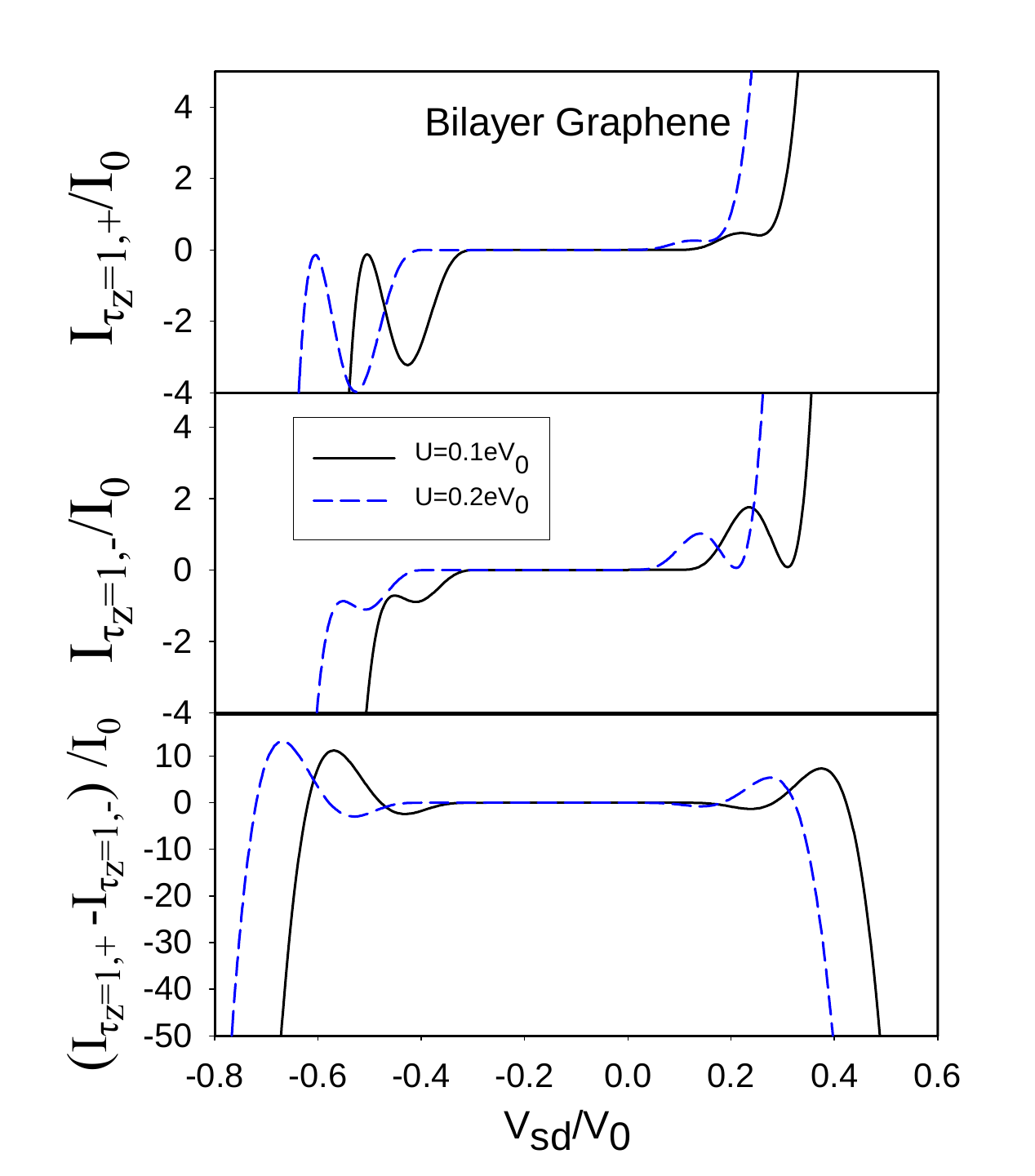}
 \caption{(Color online) I$_{\tau _z=1, \pm}$-V characteristics of bilayer graphene,
 for two values of the built-in potential. Panels (a) and (b) correspond to current flowing in the
 positive and negative
 $\hat y$-direction, respectively. In panel (c) we plot the excess of current in the positive
 $\hat y$-direction for electrons in the valley $\tau_z =1$.
 Units are $I_0= \frac {e^2}h V_0 W/d_M \times 10^5$ and $eV_0 = M_0$.}
  \label{I_Vsd_bilayer}
\end{figure}

\begin{figure}
 \includegraphics[clip,width=8cm]{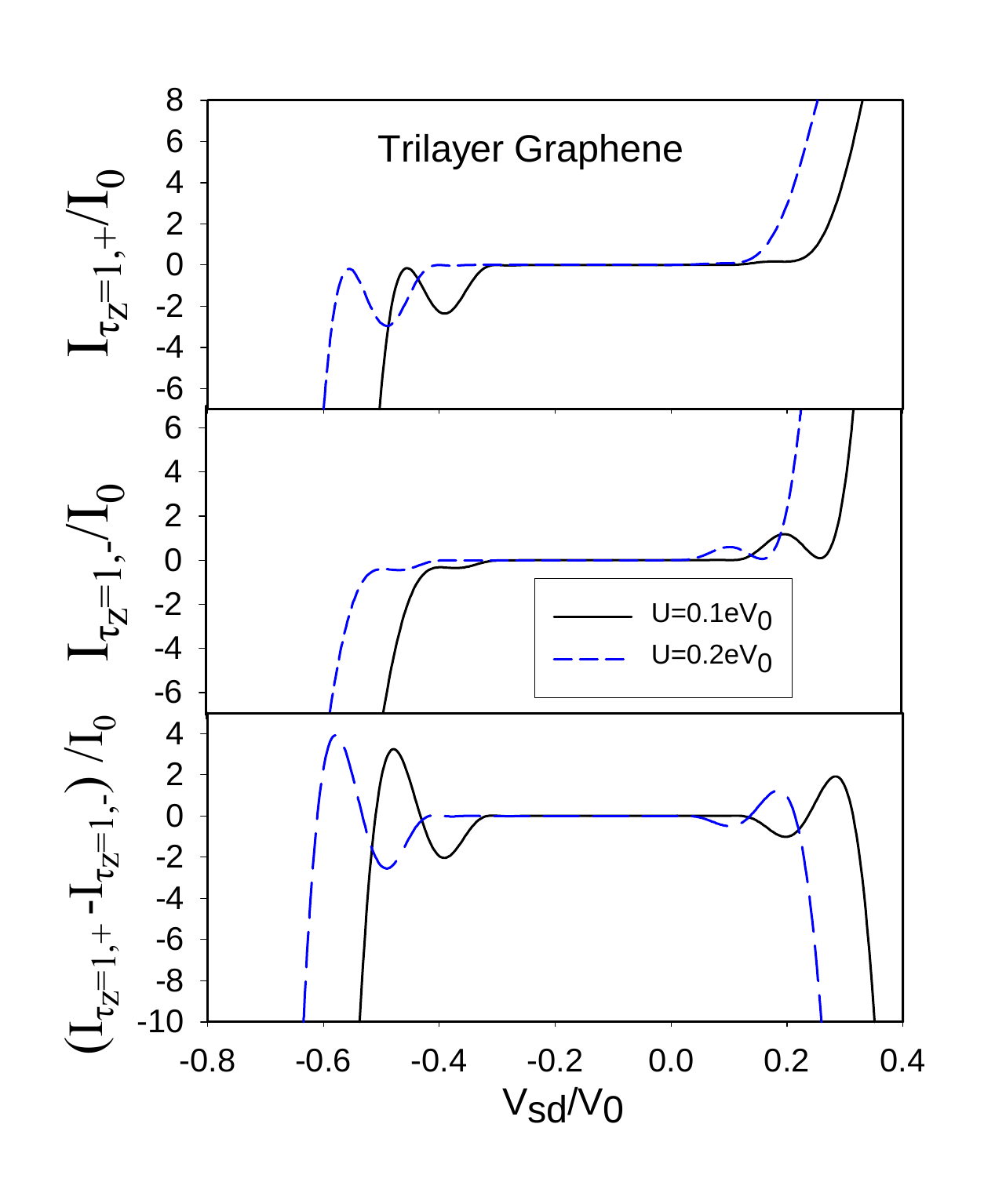}
 \caption{(Color online) I$_{\tau _z=1, \pm}$-V characteristics of trilayer graphene,
 for two values of the built-in potential. Panels (a) and (b) correspond to current flowing in the
 positive and negative
 $\hat y$-direction, respectively. In panel (c) we plot the excess of current in the positive
 $\hat y$-direction for electrons in the valley $\tau_z =1$.
 Units are $I_0= \frac {e^2}h V_0 W/d_M \times 10^4$ and $eV_0 = M_0$.}
  \label{I_Vsd_trilayer}
\end{figure}

\section{Zener tunneling current}
Finally, we calculate the tunneling current flowing through the
Zener diode. We consider a $p$-$n$ junction as the one sketched in
the inset of Fig. \ref{I_Vsd_monolayer}. $U$ represents the built-in
potential induced by doping or electrical gates, and $V_{sd}$ is the
source-drain potential difference. $L$ is the junction length. Within
the Landauer approximation, the tunneling current for index $s$
moving in the positive $\hat{y}$-direction has the form
\begin{eqnarray}
I _{s, + } &  =  &   \frac e h   \int _{-\infty} ^{\infty}   d E (
n_{E-\frac 1 2 eV_{sd}}  -   n_{E+\frac 1 2 eV_{sd}}) \nonumber \\ &
\times &\frac W {2 \pi} \int _{0} ^{q_y} T[k_y,s,\frac e L (
U+V_{sd})]d k_y, \, \, \label{I_V}
\end{eqnarray}
where $n_E$ is the Fermi-Dirac distribution and $W$ is the
transverse length of the $p$-$n$ interface. Although in the uniform
field approximation the transition amplitude is energy independent,
the limits of the integral in $k_y$ depend on  energy through the
relation $E=\varepsilon (k_x=0,q_y)$. The current flowing in the
negative $\hat y$-direction, $I _{s, - }$, is obtained performing
the integral in $k_y$ from $-q_y$ to $0$. As the transition is
dominated by small values of $k_y$, see Fig. \ref{Tvsky} and
Ref. \onlinecite{Nandkishore_2011}, we approximate $q _y $ by $\infty$
in the calculation of the currents. For zero temperature the current
gets the form
\begin{equation}
I _{s,\pm}   = \pm    \frac {e^2} h  V_{sd} \frac W {2 \pi}  \int _{
0 } ^{\pm \infty} T[k_y,s,\frac e L ( U+V_{sd})]d k_y .
\label{I_V_1}
\end{equation}
This current verifies the symmetries $I_{s,+}$=$I_{-s,-}$ and
$I_{s,-}$=$I_{-s,+}$.

In Fig. \ref{I_Vsd_monolayer} we plot the $I_{\tau _z,\pm 1}$-$V$ characteristics for
monolayer graphene and  different values of $U$. The curves present
the breakdown-type behavior characteristic of a Zener diode. In the
case of monolayer graphene the transmission amplitude is symmetric
with respect to the momentum $k_y$ and the current is equal for
positive and negative $\hat{y}$-direction.

In Fig. \ref{I_Vsd_TI} we plot the $I_{\uparrow,\pm 1}$-$V$ curves for a HgTe
quantum well  and different values of $U$. For spin up there is an
excess of current in the negative $\hat y$-direction. This effect is
the opposite for spin down electrons. These results
indicate the existence of a spin current perpendicular to the Zener
barrier. This Zener tunneling spin Hall effect is a consequence of the asymmetry in
the transition curves of Fig. \ref{Tvsky}. From Fig. \ref{I_Vsd_TI}
we obtain that  the Hall spin current can be as large as 30 per cent
of the electrical current. Because a HgTe quantum well may be a  two-dimensional topological insulator under certain conditions,  there is an extra contribution
to the current in this case coming from the spin-polarized edge states developed
in the barrier region. However, its magnitude is always of the order
of one conductance quantum $\sim e^2/h$ or less, since the electric
field diminishes it\cite{Dora_2011}. On the other hand, the Zener
tunneling spin Hall current is proportional to the transverse length
$W$, see Eq. (\ref{I_V_1}), and increases with the electric field.

In Figs. \ref{I_Vsd_bilayer} and \ref{I_Vsd_trilayer} we plot the $I$-$V$ characteristic curves for bilayer and trilayer based
Zener diodes. In both cases there are  some oscillations on top of the non-lineal, N-shaped
$I$-$V$ curves.
These negative differential conductivities appear for positive and negative $\hat y$-directions, and they have their origin in the interference between decaying states in the energy gap region\cite{Nandkishore_2011}. In trilayer graphene the negative differential conductivity is even stronger than in bilayer graphene.
At large source-to-drain voltage, the asymmetry of the tunneling amplitude is reflected in an
excess of current in the negative $\hat y$-direction with respect to the positive $\hat y$-direction.
For multilayer graphene, this effect is the opposite depending on the valley $\tau_z$. These results
indicate the existence of a valley current perpendicular to the Zener
barrier that is a consequence of the asymmetry in
the transition curves of Fig. \ref{Tvsky}.

\section{Summary}
We have analyzed Zener diode physics in HgTe quantum wells and multilayer graphene.
In the case of HgTe quantum wells we find that, after traversing the barrier, a Zener tunneling spin Hall current is developed to the right of the diode. In the case of bilayer and trilayer graphene the Zener diode generates a valley Hall current. This effect is absent for the monolayer graphene.
The magnitude and polarization of the Hall currents increase with the applied electric field.

The tunneling current is obtained from the
transmission probability that is computed numerically in the constant electric field approximation. The origin of the Hall currents is the asymmetry of the transmission probability in the momentum $k_y$
perpendicular to the tunneling barrier.
We have developed an analytical approximation for the tunneling transmission at small $k_y$ that agrees rather well with the numerical results.
The physical origin  of the Zener tunneling asymmetry  on $k_y$ is related to the Berry phase that  the carriers acquire when they are adiabatically reflected from the tunneling region.

In the case of multilayer graphene the Zener tunneling valley Hall effect could be used  for valleytronic applications\cite{Rycerz_2007,Xiao_2007}. In an appropriated geometry, the asymmetry in the Zener tunneling should enable to spatially separate the carriers of each valley\cite{Schomerus_2010}, which could be useful to manipulate the valley degree of freedom in bulk garphene.

The Zener tunneling spin Hall effect we predict to occur in HgTe quantum wells could be used for electrical manipulation of the spin currents.  The spin currents in the Zener device should be stronger than those occurring in diffusive systems, and they could be detected in
non local electrical measurements in H-shaped structures\cite{Brune_2010}.

\acknowledgments

We acknowledge fruitful discussions with B. D\'ora, S. Kohler and M. O. Goerbig. Funding for this work was provided by MICINN-Spain via grant FIS2009-08744, the CSIC JAE-Doc program, and was supported in part by the National Science Foundation under Grant No. NSF PHY05-51164.

\appendix
\section{}
In this appendix we evaluate the integral of Eq. (\ref{I_b}).
In the limit $y \rightarrow 0$, the transition probability takes the form
\begin{equation}
t =\int _{-\infty} ^{\infty} dx e ^{-i  2 \frac { \omega (x,y=0)} {\mathcal{E}} }\left ( \frac x {1+x^4}
+ i \frac {3y}{\sqrt{1+ x^4}} \right ).
\end{equation}
In terms of $\omega (x,y=0)$, the integral has simple poles in the complex plane at
$\omega _i = \int _0 ^{x_i} dx \sqrt{1+ x ^4}$, where $x_i = \pm e ^{ \pm i \pi/4}$. Expanding the value of $\omega$ near $x_i$ we find
\begin{equation}
\omega(x,y=0) - \omega _i \simeq \frac 4 3 x _i ^{3/2} ( x -x_i) ^{3/2}
\end{equation}
and we rewrite
\begin{eqnarray}
t =\! \! \!  \int _{-\infty} ^{\infty} d \omega e ^{ -i 2 \frac {\omega}{\mathcal E}} \left ( \frac {x(\omega)}{(d \omega /d x ) ^3 }+  \frac {3 i y } {(d \omega / dx )^2} \right ) \nonumber \\
= \! \!  \! \!  \sum _i  \! \! \int  _{-\infty} ^{\infty} \! \!  \! \! d \omega e ^{ -i 2 \frac {\omega}{\mathcal E}} \! \!
\left ( \frac {x(\omega)}{6 (\omega  \! - \! \omega _i) x_i ^3} \! + \! \frac {3 i y}  {6 ^{2/3}(\omega \!  - \! \omega _i)^{2/3} x_i ^2} \right ).
\end{eqnarray}
We solve this integral by closing the path around the lower half of the complex plane. This path encloses the poles $\sqrt{2}/2(\pm 1,-i)$ and their associated branches. The integral then yields
\begin{equation}
t(y)=i \frac {2 \pi} 3 e ^{- c _1 \frac 1 {\mathcal E}} \sin  {( c_ 1 \frac 1 {\mathcal E}  )}
(1-  c_2 {\mathcal E} ^{1/3} y),
\end{equation}
with $c_1=\frac 1 4 \sqrt{\frac {\pi} 2} \frac {\Gamma (1/4)}{\Gamma(7/4)}$ and $c_2 =\frac { 3 ^{4/3}}   {\Gamma (2/3)}$.

\bibliography{TI}

\begin{thebibliography}{44}
\expandafter\ifx\csname natexlab\endcsname\relax\def\natexlab#1{#1}\fi
\expandafter\ifx\csname bibnamefont\endcsname\relax
  \def\bibnamefont#1{#1}\fi
\expandafter\ifx\csname bibfnamefont\endcsname\relax
  \def\bibfnamefont#1{#1}\fi
\expandafter\ifx\csname citenamefont\endcsname\relax
  \def\citenamefont#1{#1}\fi
\expandafter\ifx\csname url\endcsname\relax
  \def\url#1{\texttt{#1}}\fi
\expandafter\ifx\csname urlprefix\endcsname\relax\def\urlprefix{URL }\fi
\providecommand{\bibinfo}[2]{#2}
\providecommand{\eprint}[2][]{\url{#2}}

\bibitem[{\citenamefont{Sze}(1981)}]{Sze_book}
\bibinfo{author}{\bibfnamefont{S.~M.} \bibnamefont{Sze}},
  \emph{\bibinfo{title}{Physics of Semiconductor Devices}}
  (\bibinfo{publisher}{Wiley}, \bibinfo{year}{1981}).

\bibitem[{\citenamefont{Zener}(1934)}]{Zener_1934}
\bibinfo{author}{\bibfnamefont{C.}~\bibnamefont{Zener}},
  \bibinfo{journal}{Proc. Royal Soc. London} p. \bibinfo{pages}{523}
  (\bibinfo{year}{1934}).

\bibitem[{\citenamefont{Wittig}(2005)}]{Wittig_2005}
\bibinfo{author}{\bibfnamefont{C.}~\bibnamefont{Wittig}}, \bibinfo{journal}{J.
  Phys. Chem. B} \textbf{\bibinfo{volume}{109}}, \bibinfo{pages}{8428}
  (\bibinfo{year}{2005}).

\bibitem[{\citenamefont{Kane and Blount}(1969)}]{Kane_Blount}
\bibinfo{author}{\bibfnamefont{E.~O.} \bibnamefont{Kane}} \bibnamefont{and}
  \bibinfo{author}{\bibfnamefont{E.}~\bibnamefont{Blount}},
  \emph{\bibinfo{title}{Tunneling Phenomena in Solids}}
  (\bibinfo{publisher}{Plenum Press,, New York}, \bibinfo{year}{1969}).

\bibitem[{\citenamefont{Shevchenko et~al.}(2010)\citenamefont{Shevchenko,
  Adshab, and Nori}}]{Shevchenko_2010}
\bibinfo{author}{\bibfnamefont{S.~N.} \bibnamefont{Shevchenko}},
  \bibinfo{author}{\bibfnamefont{S.}~\bibnamefont{Adshab}}, \bibnamefont{and}
  \bibinfo{author}{\bibfnamefont{F.}~\bibnamefont{Nori}},
  \bibinfo{journal}{Physics Reports} \textbf{\bibinfo{volume}{492}},
  \bibinfo{pages}{1} (\bibinfo{year}{2010}).

\bibitem[{\citenamefont{Shimshoni and Gefel}(1991)}]{Shimshoni_1991}
\bibinfo{author}{\bibfnamefont{E.}~\bibnamefont{Shimshoni}} \bibnamefont{and}
  \bibinfo{author}{\bibfnamefont{Y.}~\bibnamefont{Gefel}},
  \bibinfo{journal}{Annals Phys.} \textbf{\bibinfo{volume}{201}},
  \bibinfo{pages}{16} (\bibinfo{year}{1991}).

\bibitem[{\citenamefont{{ X.-L. Qi {\it et al.}}}(2006)}]{Qi_2006}
\bibinfo{author}{\bibnamefont{{ X.-L. Qi {\it et al.}}}},
  \bibinfo{journal}{Phys. Rev. B} \textbf{\bibinfo{volume}{74}},
  \bibinfo{pages}{085308} (\bibinfo{year}{2006}).

\bibitem[{\citenamefont{{ B. Bernevig {\it et al.}}}(2006)}]{Bernevig_2006}
\bibinfo{author}{\bibnamefont{{ B. Bernevig {\it et al.}}}},
  \bibinfo{journal}{Science} \textbf{\bibinfo{volume}{314}},
  \bibinfo{pages}{1757} (\bibinfo{year}{2006}).

\bibitem[{\citenamefont{{ M. K\"onig {\it et al.}}}(2007)}]{Konig_2007}
\bibinfo{author}{\bibnamefont{{ M. K\"onig {\it et al.}}}},
  \bibinfo{journal}{Science} \textbf{\bibinfo{volume}{318}},
  \bibinfo{pages}{766} (\bibinfo{year}{2007}).

\bibitem[{\citenamefont{{ L. Fu {\it et al.}}}(2007)}]{Fu_2007}
\bibinfo{author}{\bibnamefont{{ L. Fu {\it et al.}}}}, \bibinfo{journal}{Phys.
  Rev. Lett.} \textbf{\bibinfo{volume}{98}}, \bibinfo{pages}{106803}
  (\bibinfo{year}{2007}).

\bibitem[{\citenamefont{{ J. E. Moore {\it etal.}}}(2007)}]{Moore_2007}
\bibinfo{author}{\bibnamefont{{ J. E. Moore {\it etal.}}}},
  \bibinfo{journal}{Phys. Rev. B} \textbf{\bibinfo{volume}{75}},
  \bibinfo{pages}{121306} (\bibinfo{year}{2007}).

\bibitem[{\citenamefont{Murakami}(2007)}]{Murakami_2007}
\bibinfo{author}{\bibfnamefont{S.}~\bibnamefont{Murakami}},
  \bibinfo{journal}{New Journal of Physics} \textbf{\bibinfo{volume}{9}},
  \bibinfo{pages}{356} (\bibinfo{year}{2007}).

\bibitem[{\citenamefont{Novoselov et~al.}(2004)\citenamefont{Novoselov, Geim,
  Mozorov, Jiang, Zhang, Dubonos, Gregorieva, and Firsov}}]{Novoselov_2004}
\bibinfo{author}{\bibfnamefont{K.~S.} \bibnamefont{Novoselov}},
  \bibinfo{author}{\bibfnamefont{A.~K.} \bibnamefont{Geim}},
  \bibinfo{author}{\bibfnamefont{S.~V.} \bibnamefont{Mozorov}},
  \bibinfo{author}{\bibfnamefont{D.}~\bibnamefont{Jiang}},
  \bibinfo{author}{\bibfnamefont{Y.}~\bibnamefont{Zhang}},
  \bibinfo{author}{\bibfnamefont{S.~V.} \bibnamefont{Dubonos}},
  \bibinfo{author}{\bibfnamefont{I.~V.} \bibnamefont{Gregorieva}},
  \bibnamefont{and} \bibinfo{author}{\bibfnamefont{A.~A.}
  \bibnamefont{Firsov}}, \bibinfo{journal}{Science}
  \textbf{\bibinfo{volume}{306}}, \bibinfo{pages}{666} (\bibinfo{year}{2004}).

\bibitem[{\citenamefont{Novoselov et~al.}(2005)\citenamefont{Novoselov, Jiang,
  Booth, Khotkevich, Morozov, and Geim}}]{Novoselov_2005}
\bibinfo{author}{\bibfnamefont{K.~S.} \bibnamefont{Novoselov}},
  \bibinfo{author}{\bibfnamefont{D.}~\bibnamefont{Jiang}},
  \bibinfo{author}{\bibfnamefont{T.}~\bibnamefont{Booth}},
  \bibinfo{author}{\bibfnamefont{V.~V.} \bibnamefont{Khotkevich}},
  \bibinfo{author}{\bibfnamefont{S.~M.} \bibnamefont{Morozov}},
  \bibnamefont{and} \bibinfo{author}{\bibfnamefont{A.~K.} \bibnamefont{Geim}},
  \bibinfo{journal}{Nature} \textbf{\bibinfo{volume}{438}},
  \bibinfo{pages}{197} (\bibinfo{year}{2005}).

\bibitem[{\citenamefont{Zhang et~al.}(2005)\citenamefont{Zhang, Tan, Stormer,
  and Kim}}]{Zhang_2005}
\bibinfo{author}{\bibfnamefont{Y.}~\bibnamefont{Zhang}},
  \bibinfo{author}{\bibfnamefont{Y.-W.} \bibnamefont{Tan}},
  \bibinfo{author}{\bibfnamefont{H.~L.} \bibnamefont{Stormer}},
  \bibnamefont{and} \bibinfo{author}{\bibfnamefont{P.}~\bibnamefont{Kim}},
  \bibinfo{journal}{Nature} \textbf{\bibinfo{volume}{438}},
  \bibinfo{pages}{201} (\bibinfo{year}{2005}).

\bibitem[{\citenamefont{Katsnelson et~al.}(2006)\citenamefont{Katsnelson,
  Novoselov, and Geim}}]{Katsnelson_2006a}
\bibinfo{author}{\bibfnamefont{M.~I.} \bibnamefont{Katsnelson}},
  \bibinfo{author}{\bibfnamefont{K.~S.} \bibnamefont{Novoselov}},
  \bibnamefont{and} \bibinfo{author}{\bibfnamefont{A.}~\bibnamefont{Geim}},
  \bibinfo{journal}{Nat. Phys.} \textbf{\bibinfo{volume}{2}},
  \bibinfo{pages}{620} (\bibinfo{year}{2006}).

\bibitem[{\citenamefont{Cheianov and Fal'ko}(2006)}]{Cheianov_2006}
\bibinfo{author}{\bibfnamefont{V.~V.} \bibnamefont{Cheianov}} \bibnamefont{and}
  \bibinfo{author}{\bibfnamefont{V.~I.} \bibnamefont{Fal'ko}},
  \bibinfo{journal}{Phys. Rev. B} \textbf{\bibinfo{volume}{74}},
  \bibinfo{pages}{041403} (\bibinfo{year}{2006}).

\bibitem[{\citenamefont{Williams et~al.}(2007)\citenamefont{Williams, DiCarlo,
  and Marcus}}]{Williams_2007}
\bibinfo{author}{\bibfnamefont{J.~R.} \bibnamefont{Williams}},
  \bibinfo{author}{\bibfnamefont{L.}~\bibnamefont{DiCarlo}}, \bibnamefont{and}
  \bibinfo{author}{\bibfnamefont{C.~M.} \bibnamefont{Marcus}},
  \bibinfo{journal}{Science} \textbf{\bibinfo{volume}{317}},
  \bibinfo{pages}{638} (\bibinfo{year}{2007}).

\bibitem[{\citenamefont{Zhang and Fogler}(2008)}]{Zhang_2008}
\bibinfo{author}{\bibfnamefont{L.~M.} \bibnamefont{Zhang}} \bibnamefont{and}
  \bibinfo{author}{\bibfnamefont{M.~M.} \bibnamefont{Fogler}},
  \bibinfo{journal}{Phys. Rev. Lett.} \textbf{\bibinfo{volume}{100}},
  \bibinfo{pages}{116804} (\bibinfo{year}{2008}).

\bibitem[{\citenamefont{Young and Kim}(2009)}]{Young_2009}
\bibinfo{author}{\bibfnamefont{A.}~\bibnamefont{Young}} \bibnamefont{and}
  \bibinfo{author}{\bibfnamefont{P.}~\bibnamefont{Kim}}, \bibinfo{journal}{Nat.
  Phys.} \textbf{\bibinfo{volume}{5}}, \bibinfo{pages}{222}
  (\bibinfo{year}{2009}).

\bibitem[{\citenamefont{Vandecasteele et~al.}(2010)\citenamefont{Vandecasteele,
  Barreiro, Lazzeri, Bachtold, and Mauri}}]{Vandecasteele_2010}
\bibinfo{author}{\bibfnamefont{N.}~\bibnamefont{Vandecasteele}},
  \bibinfo{author}{\bibfnamefont{A.}~\bibnamefont{Barreiro}},
  \bibinfo{author}{\bibfnamefont{M.}~\bibnamefont{Lazzeri}},
  \bibinfo{author}{\bibfnamefont{A.}~\bibnamefont{Bachtold}}, \bibnamefont{and}
  \bibinfo{author}{\bibfnamefont{F.}~\bibnamefont{Mauri}},
  \bibinfo{journal}{Phys. Rev. B} \textbf{\bibinfo{volume}{82}},
  \bibinfo{pages}{045416} (\bibinfo{year}{2010}).

\bibitem[{\citenamefont{Stander et~al.}(2009)\citenamefont{Stander, Huard, and
  Goldhaber-Gordon}}]{Stander_2009}
\bibinfo{author}{\bibfnamefont{N.}~\bibnamefont{Stander}},
  \bibinfo{author}{\bibfnamefont{B.}~\bibnamefont{Huard}}, \bibnamefont{and}
  \bibinfo{author}{\bibfnamefont{D.}~\bibnamefont{Goldhaber-Gordon}},
  \bibinfo{journal}{Phys. Rev. Lett.} \textbf{\bibinfo{volume}{102}},
  \bibinfo{pages}{026807} (\bibinfo{year}{2009}).

\bibitem[{\citenamefont{Jena et~al.}(2008)\citenamefont{Jena, Fang, Zhang, and
  Xing}}]{Jena_2008}
\bibinfo{author}{\bibfnamefont{D.}~\bibnamefont{Jena}},
  \bibinfo{author}{\bibfnamefont{T.}~\bibnamefont{Fang}},
  \bibinfo{author}{\bibfnamefont{Q.}~\bibnamefont{Zhang}}, \bibnamefont{and}
  \bibinfo{author}{\bibfnamefont{H.}~\bibnamefont{Xing}},
  \bibinfo{journal}{Applied Physics Letters} \textbf{\bibinfo{volume}{93}},
  \bibinfo{pages}{112106} (\bibinfo{year}{2008}).

\bibitem[{\citenamefont{Chiu et~al.}(2010)\citenamefont{Chiu, Perebeinos, Lin,
  and Avouris}}]{Chiu_2010}
\bibinfo{author}{\bibfnamefont{H.-Y.} \bibnamefont{Chiu}},
  \bibinfo{author}{\bibfnamefont{V.}~\bibnamefont{Perebeinos}},
  \bibinfo{author}{\bibfnamefont{Y.-M.} \bibnamefont{Lin}}, \bibnamefont{and}
  \bibinfo{author}{\bibfnamefont{P.}~\bibnamefont{Avouris}},
  \bibinfo{journal}{Nano Letters} \textbf{\bibinfo{volume}{10}},
  \bibinfo{pages}{4634} (\bibinfo{year}{2010}).

\bibitem[{\citenamefont{Brey and Fertig}(2009)}]{Brey_2009}
\bibinfo{author}{\bibfnamefont{L.}~\bibnamefont{Brey}} \bibnamefont{and}
  \bibinfo{author}{\bibfnamefont{H.~A.} \bibnamefont{Fertig}},
  \bibinfo{journal}{Phys. Rev. Lett.} \textbf{\bibinfo{volume}{103}},
  \bibinfo{pages}{046809} (\bibinfo{year}{2009}).

\bibitem[{\citenamefont{Arovas et~al.}(2010)\citenamefont{Arovas, Brey, Fertig,
  Kim, and Ziegler}}]{Arovas_2010}
\bibinfo{author}{\bibfnamefont{D.~P.} \bibnamefont{Arovas}},
  \bibinfo{author}{\bibfnamefont{L.}~\bibnamefont{Brey}},
  \bibinfo{author}{\bibfnamefont{H.~A.} \bibnamefont{Fertig}},
  \bibinfo{author}{\bibfnamefont{E.-A.} \bibnamefont{Kim}}, \bibnamefont{and}
  \bibinfo{author}{\bibfnamefont{K.}~\bibnamefont{Ziegler}},
  \bibinfo{journal}{New Journal of Physics} \textbf{\bibinfo{volume}{12}},
  \bibinfo{pages}{123020} (\bibinfo{year}{2010}).

\bibitem[{\citenamefont{Nandkishore and Levitov}(2011)}]{Nandkishore_2011}
\bibinfo{author}{\bibfnamefont{R.}~\bibnamefont{Nandkishore}} \bibnamefont{and}
  \bibinfo{author}{\bibfnamefont{L.}~\bibnamefont{Levitov}},
  \bibinfo{journal}{Proceedings of the National Academy of Sciences}
  \textbf{\bibinfo{volume}{108}}, \bibinfo{pages}{14021}
  (\bibinfo{year}{2011}).

\bibitem[{par()}]{parameters}
\bibinfo{note}{Typical parameters for HgTe quantum wells
  are\cite{Tkachov_2011}, $M_0 \! \sim \! \!10\mathrm{meV}$, $B\! \! \sim \!
  \!1.2eV \cdot \mathrm{nm}^2$, $A \! \!\sim \! \!0.38\mathrm{eV} \cdot
  \mathrm{nm}$.}

\bibitem[{\citenamefont{Kane and Mele}(2005{\natexlab{a}})}]{Kane_2005a}
\bibinfo{author}{\bibfnamefont{C.~L.} \bibnamefont{Kane}} \bibnamefont{and}
  \bibinfo{author}{\bibfnamefont{E.~J.} \bibnamefont{Mele}},
  \bibinfo{journal}{Phys. Rev. Lett.} \textbf{\bibinfo{volume}{95}},
  \bibinfo{pages}{226801} (\bibinfo{year}{2005}{\natexlab{a}}).

\bibitem[{\citenamefont{Kane and Mele}(2005{\natexlab{b}})}]{Kane_2005b}
\bibinfo{author}{\bibfnamefont{C.~L.} \bibnamefont{Kane}} \bibnamefont{and}
  \bibinfo{author}{\bibfnamefont{E.~J.} \bibnamefont{Mele}},
  \bibinfo{journal}{Phys. Rev. Lett.} \textbf{\bibinfo{volume}{95}},
  \bibinfo{pages}{146802} (\bibinfo{year}{2005}{\natexlab{b}}).

\bibitem[{\citenamefont{Prada et~al.}(2011)\citenamefont{Prada, San-Jose, Brey,
  and Fertig}}]{Prada_2011}
\bibinfo{author}{\bibfnamefont{E.}~\bibnamefont{Prada}},
  \bibinfo{author}{\bibfnamefont{P.}~\bibnamefont{San-Jose}},
  \bibinfo{author}{\bibfnamefont{L.}~\bibnamefont{Brey}}, \bibnamefont{and}
  \bibinfo{author}{\bibfnamefont{H.}~\bibnamefont{Fertig}},
  \bibinfo{journal}{Solid State Communications} \textbf{\bibinfo{volume}{151}},
  \bibinfo{pages}{1075 } (\bibinfo{year}{2011}).

\bibitem[{\citenamefont{Castro-Neto et~al.}(2009)\citenamefont{Castro-Neto,
  F.Guinea, N.M.R.Peres, K.S.Novoselov, and A.K.Geim}}]{Castro_Neto_RMP}
\bibinfo{author}{\bibfnamefont{A.~H.} \bibnamefont{Castro-Neto}},
  \bibinfo{author}{\bibnamefont{F.Guinea}},
  \bibinfo{author}{\bibnamefont{N.M.R.Peres}},
  \bibinfo{author}{\bibnamefont{K.S.Novoselov}}, \bibnamefont{and}
  \bibinfo{author}{\bibnamefont{A.K.Geim}}, \bibinfo{journal}{Rev.\ Mod.\
  Phys.} \textbf{\bibinfo{volume}{81}}, \bibinfo{pages}{109}
  (\bibinfo{year}{2009}).

\bibitem[{\citenamefont{Min et~al.}(2008)\citenamefont{Min, Borghi, Polini, and
  MacDonald}}]{Min_2008}
\bibinfo{author}{\bibfnamefont{H.}~\bibnamefont{Min}},
  \bibinfo{author}{\bibfnamefont{G.}~\bibnamefont{Borghi}},
  \bibinfo{author}{\bibfnamefont{M.}~\bibnamefont{Polini}}, \bibnamefont{and}
  \bibinfo{author}{\bibfnamefont{A.~H.} \bibnamefont{MacDonald}},
  \bibinfo{journal}{Phys. Rev. B} \textbf{\bibinfo{volume}{77}},
  \bibinfo{pages}{041407} (\bibinfo{year}{2008}).

\bibitem[{\citenamefont{McCann et~al.}(2007)\citenamefont{McCann, Abergel, and
  Fal'ko}}]{McCann_SSC}
\bibinfo{author}{\bibfnamefont{E.}~\bibnamefont{McCann}},
  \bibinfo{author}{\bibfnamefont{D.~S.} \bibnamefont{Abergel}},
  \bibnamefont{and} \bibinfo{author}{\bibfnamefont{V.~I.}
  \bibnamefont{Fal'ko}}, \bibinfo{journal}{Solid State Communications}
  \textbf{\bibinfo{volume}{143}}, \bibinfo{pages}{110 } (\bibinfo{year}{2007}).

\bibitem[{\citenamefont{Hasan and Kane}(2010)}]{hasan_2010}
\bibinfo{author}{\bibfnamefont{M.~Z.} \bibnamefont{Hasan}} \bibnamefont{and}
  \bibinfo{author}{\bibfnamefont{C.~L.} \bibnamefont{Kane}},
  \bibinfo{journal}{Rev. Mod. Phys.} \textbf{\bibinfo{volume}{82}},
  \bibinfo{pages}{3045} (\bibinfo{year}{2010}).

\bibitem[{\citenamefont{Qi and Zhang}(2011)}]{Qi_2011}
\bibinfo{author}{\bibfnamefont{X.-L.} \bibnamefont{Qi}} \bibnamefont{and}
  \bibinfo{author}{\bibfnamefont{S.-C.} \bibnamefont{Zhang}},
  \bibinfo{journal}{Rev. Mod. Phys.} \textbf{\bibinfo{volume}{83}},
  \bibinfo{pages}{1057} (\bibinfo{year}{2011}).

\bibitem[{\citenamefont{Guigou et~al.}(2011)\citenamefont{Guigou, Recher,
  Cayssol, and Trauzettel}}]{Guigou_2011}
\bibinfo{author}{\bibfnamefont{M.}~\bibnamefont{Guigou}},
  \bibinfo{author}{\bibfnamefont{P.}~\bibnamefont{Recher}},
  \bibinfo{author}{\bibfnamefont{J.}~\bibnamefont{Cayssol}}, \bibnamefont{and}
  \bibinfo{author}{\bibfnamefont{B.}~\bibnamefont{Trauzettel}},
  \bibinfo{journal}{Phys. Rev. B} \textbf{\bibinfo{volume}{84}},
  \bibinfo{pages}{094534} (\bibinfo{year}{2011}).

\bibitem[{\citenamefont{Davis and Pechukas}(1976)}]{Davis_1976}
\bibinfo{author}{\bibfnamefont{J.}~\bibnamefont{Davis}} \bibnamefont{and}
  \bibinfo{author}{\bibfnamefont{P.}~\bibnamefont{Pechukas}},
  \bibinfo{journal}{J.Chem.Phys.} \textbf{\bibinfo{volume}{64}},
  \bibinfo{pages}{3129} (\bibinfo{year}{1976}).

\bibitem[{\citenamefont{D\'ora and Moessner}(2011)}]{Dora_2011}
\bibinfo{author}{\bibfnamefont{B.}~\bibnamefont{D\'ora}} \bibnamefont{and}
  \bibinfo{author}{\bibfnamefont{R.}~\bibnamefont{Moessner}},
  \bibinfo{journal}{Phys. Rev. B} \textbf{\bibinfo{volume}{83}},
  \bibinfo{pages}{073403} (\bibinfo{year}{2011}).

\bibitem[{\citenamefont{Rycerz et~al.}(2007)\citenamefont{Rycerz, Tworzydlo,
  and Beenakker}}]{Rycerz_2007}
\bibinfo{author}{\bibfnamefont{A.}~\bibnamefont{Rycerz}},
  \bibinfo{author}{\bibfnamefont{J.}~\bibnamefont{Tworzydlo}},
  \bibnamefont{and} \bibinfo{author}{\bibfnamefont{C.~W.~J.}
  \bibnamefont{Beenakker}}, \bibinfo{journal}{Nature Physics}
  \textbf{\bibinfo{volume}{3}}, \bibinfo{pages}{172} (\bibinfo{year}{2007}).

\bibitem[{\citenamefont{Xiao et~al.}(2007)\citenamefont{Xiao, Yao, and
  Niu}}]{Xiao_2007}
\bibinfo{author}{\bibfnamefont{D.}~\bibnamefont{Xiao}},
  \bibinfo{author}{\bibfnamefont{W.}~\bibnamefont{Yao}}, \bibnamefont{and}
  \bibinfo{author}{\bibfnamefont{Q.}~\bibnamefont{Niu}},
  \bibinfo{journal}{Phys. Rev. Lett.} \textbf{\bibinfo{volume}{99}},
  \bibinfo{pages}{236809} (\bibinfo{year}{2007}).

\bibitem[{\citenamefont{Schomerus}(2010)}]{Schomerus_2010}
\bibinfo{author}{\bibfnamefont{H.}~\bibnamefont{Schomerus}},
  \bibinfo{journal}{Phys. Rev. B} \textbf{\bibinfo{volume}{82}},
  \bibinfo{pages}{165409} (\bibinfo{year}{2010}).

\bibitem[{\citenamefont{{C. Brune {\it et al.}}}(2010)}]{Brune_2010}
\bibinfo{author}{\bibnamefont{{C. Brune {\it et al.}}}},
  \bibinfo{journal}{Nat.Phys.} \textbf{\bibinfo{volume}{6}},
  \bibinfo{pages}{448} (\bibinfo{year}{2010}).

\bibitem[{\citenamefont{{G. Tkachov {\it et al.}}}(2011)}]{Tkachov_2011}
\bibinfo{author}{\bibnamefont{{G. Tkachov {\it et al.}}}},
  \bibinfo{journal}{Phys. Rev. Lett.} \textbf{\bibinfo{volume}{106}},
  \bibinfo{pages}{076802} (\bibinfo{year}{2011}).

\end{thebibliography}

\end{document}